\documentclass[a4paper,11pt]{article}

\usepackage[margin=1in]{geometry}
\usepackage{authblk}
\usepackage{subcaption}

\usepackage[table,xcdraw]{xcolor}
\usepackage[normalem]{ulem}
\usepackage{amsmath}
\usepackage{amssymb}
\usepackage{amsthm}
\usepackage{graphicx}
\usepackage{epstopdf}
\usepackage{caption}
\usepackage[ruled,vlined]{algorithm2e}
\usepackage{bm} 
\usepackage{url} 
\usepackage{multirow}
\usepackage{tabularx}
\usepackage{lineno}
\usepackage{soul}
\usepackage{amsfonts}
\usepackage{bbm}
\usepackage{float}
\usepackage{algorithmic}
\usepackage{amssymb}
\usepackage{mathtools}
\usepackage{apacite}

\newcommand{\cf}{{\it cf.}~}

\title{Forecasting Local Ionospheric Parameters Using Transformers}
\author[1,2,3]{Daniel J.  Alford-Lago\thanks{daniel.j.alford-lago.civ@us.navy.mil}}
\author[1]{Christopher W. Curtis}
\author[3]{Alexander T. Ihler}
\author[4]{Katherine A. Zawdie}
\author[4]{Douglas P. Drob}
\affil[1]{Atmospheric Propagation Branch, Naval Information Warfare Center Pacific, San Diego, California, USA}
\affil[2]{Department of Mathematics and Statistics, San Diego State University, San Diego, California, USA}
\affil[3]{Department of Computer Science, UC Irvine, Irvine, California, USA}
\affil[4]{Space Science Division, Naval Research Laboratory, Washington, District of Columbia, USA}

\date{}
\begin{document}
\maketitle

\begin{abstract}
We present a novel method for forecasting key ionospheric parameters using transformer-based neural networks. The model provides accurate forecasts and uncertainty quantification of the F2-layer peak plasma frequency (foF2), the F2-layer peak density height (hmF2), and total electron content (TEC) for a given geographic location. It includes a number of exogenous variables, including F10.7cm solar flux and disturbance storm time (Dst). We demonstrate how transformers can be trained in a data assimilation-like fashion that uses these exogenous variables along with na\"ive predictions from climatology to generate 24-hour forecasts with nonparametric uncertainty bounds. We call this method the Local Ionospheric Forecast Transformer (LIFT). We demonstrate that the trained model can generalize to new geographic locations and time periods not seen during training, and we compare its performance to that of the International Reference Ionosphere (IRI) using CCIR coefficients.
\end{abstract}

\section{Introduction}\label{sec-intro}
Accurate and efficient modeling of Earth's ionosphere has a significant impact on research and operational communities due to its effects on radio communications, radar performance \cite{budden, davies_1990, ratcliffe_1959}, and satellite drag \cite{berger_2023}. Success in forecasting key parameters such as the F2-layer critical frequency (foF2) and height (hmF2) and the total electron content (TEC) allows one to anticipate and mitigate the impacts of ionospheric variability on such systems. Over the past decades, many modeling approaches have been developed to predict these ionospheric parameters with increasing accuracy and skill. These models may be broadly categorized as empirical, physics-based, and, more recently, machine learning methods.

Empirical models often rely on extensive historical datasets to establish statistical relationships between ionospheric parameters and geophysical variables. The International Reference Ionosphere (IRI) model \cite{bilitza_2022} is a widely used standard that provides monthly averages of various ionospheric parameters based on many decades of past observations. IRI has seen continual development and improvements over the years, adding a host of submodels used to capture specific aspects of the ionosphere such as the CCIR \cite{jones_1962, jones_1965} and URSI \cite{fox_1988} foF2 models for representing the diurnal variations of the peak plasma density across the globe, the AMTB \cite{altadill_2013} and SHU-2015 \cite{shubin_2015} models for harmonic expansions of hmF2, and NeQuick 2 \cite{nava_2008} for improved topside electron density accuracy and thus better estimates of TEC \cite{bilitza_2009, kashcheyev_2019}. While large empirical models like IRI continue to improve, the number of available options needed to address each domain and source of variance in the ionosphere also grows, and choosing the appropriate settings may be prohibitive without expert knowledge of each submodel.

Physics-based models instead aim to incorporate the fundamental processes governing the dynamics of the ionosphere and its drivers by solving highly complex sets of equations related to plasma magnetohydrodynamics, electrodynamics, and chemistry. These models seek to simulate the ionospheric response to solar and geomagnetic inputs. The Thermosphere-Ionosphere-Mesosphere-Electrodynamics General Circulation Model (TIME-GCM) \cite{roble_1994}, the Coupled Thermosphere Ionosphere Plasmasphere Electrodynamics (CTIPe) model \cite{millward_2001}, and SAMI3 \cite{huba_2013} are examples of such models. While they offer detailed insights through highly accurate nowcast and hindcast, their forecast skill can drop precipitously over relatively short time windows, even with careful treatment of driving parameters such as solar and geomagnetic activity. Additionally, the simulations themselves can be computationally expensive, requiring high-performance computing to generate high-resolution or large numbers of simulations.

Both empirical and physics-based methods require special treatment of drivers and parameterizations in order to maintain accuracy over a forecast period. Although there are efforts to develop better indices for these drivers \cite{nishioka_2017}, the impact on forecast skill is largely determined by how well these drivers can inform future values of the parameters and not just how correlated they are in the past. This challenge is further complicated by the fact that these drivers, which often take the form of global indices, may interact with the forecasted parameters with unknown or varying time delays \cite{schmolter_2024, chen_2018, chen_2015, coley_2012, jakowski_1991}. A further limitation of current empirical and physics-based models is their inability to provide robust uncertainty quantification. The dual challenge of capturing complex, time-delayed driver interactions and providing reliable uncertainty estimates motivates the integration of machine learning into next-generation forecasting systems.

Many machine learning models operate on learned latent representations of the data. These learned spaces can greatly simplify a time series through linearizations of the latent variable dynamics \cite{lago_2022} or time delay embeddings \cite{curtis_2024}. However, the burden of choosing the correct latent dimension or time delay can make these methods cumbersome when applied to real-world datasets, and there is no straightforward means of validating that a given embedding will continue to work for future time series or for new geographic locations. Recent advances in transformer-based neural networks offer a promising alternative \cite{lim_2021, zhou_2023, tang_2023, zhou_2025, mao_2025}.

The use of transformers for forecasting in space-weather applications has grown significantly in recent years. Current work primarily uses recurrent networks and attention-based architectures for ionospheric forecasting, with the bulk of activity focused on global TEC maps. For instance, sequence-to-sequence transformers trained on global ionospheric maps (GIMs) from the Center for Orbit Determination in Europe (CODE) achieve stable multi-step forecasts up to 48 hours, with root-mean-squared error (RMSE) reported near 1.8 TECU \cite{shih2024}. Extensions that pair transformers with representation learning or denoising have also been developed. For example, an ``improved transformer'' that integrates deep denoising autoencoders to compress and clean GIMs reduces errors relative to conventional baselines and demonstrates better robustness during disturbed conditions \cite{Wu2024}. Other efforts exploit autocorrelation mechanisms and synthetic data augmentation to capture long-range dependencies, as seen in a synthesis-style autocorrelation transformer that provides one-day GIM forecasts and outperforms operational one-day products on a 2018 benchmark \cite{yuan2023}. Capturing spatial couplings explicitly has also proved beneficial. A graph-enabled spatiotemporal transformer (GEST) models station connectivity with graphical-model layers, while a time-transformer attention block handles temporal dependencies; this approach yields higher accuracy than prior deep models for multistep TEC prediction and improves storm-time performance \cite{yu2024}. More hybrid designs have appeared, combining convolutional encoders for local spatial patterns with transformer temporal modules, such as convolutional attentional image time-sequence transformers for TEC maps \cite{xia2022}. More recently, autoformer variants with decomposition layers have been applied to global TEC forecasting \cite{zhou_2025_edautoformer}, and deep learning methods including LSTM and GRU have been developed toward operational forecasting services with incremental learning capabilities \cite{molina_2025}. Related spatiotemporal encoder--decoder variants claim state-of-the-art results across solar-cycle phases.

These studies collectively point to several design levers that matter for TEC forecasting, including explicit spatial modeling (graphs or CNNs), long receptive fields via attention, and incorporation of drivers such as F10.7, Kp, Dst, and solar-wind parameters. For foF2, recent transformer applications adopt long-sequence variants tailored to single-station forecasting. Informer architectures \cite{zhou_2021}, which use probabilistic sparse attention and distillation to reduce complexity, improve 5- to 48-hour foF2 forecasts over recurrent (LSTM) and empirical (IRI) models at both middle and low latitudes \cite{bi2022}. Reported implementations train on multi-year ionosonde records with exogenous inputs (F10.7, Dst, Kp), maintain skill during geomagnetic storms, and reduce RMSE across the forecast horizon compared with tuned recurrent baselines \cite{qiao2024}. These results indicate that attention mechanisms help recover diurnal and seasonal structure while adapting to storm-time deviations when given driver indices. By contrast, explicit transformer formulations for hmF2 remain sparse in the open literature; most studies targeting this parameter still rely on feedforward neural networks or recurrent models and report improvements over IRI or simple autoregressive baselines \cite{tulasi2018}.

However, existing transformer models often use multiple attention layers with large embedding dimensions, which drastically increases model size relative to the available data. For instance, the Informer foF2 forecast model in \citeA{bi2022} uses 10 encoder layers and 8 decoder layers, each with embedding dimensions of 512 and feed-forward networks with a 2048 inner dimension, resulting in tens of millions of trainable parameters. This is especially relevant for spatiotemporal data such as GIMs or ionosonde time series, as batching generates highly correlated samples. Consequently, there is a gap in the existing literature for parsimonious transformer-based forecasters that jointly predict foF2, hmF2, and TEC, leverage exogenous drivers, and provide calibrated uncertainty; \cf \citeA{mao_2025} for a comprehensive review demonstrating this gap.

We address this gap with the Local Ionospheric Forecast Transformer (LIFT). Our approach differs from prior work in several respects. First, we use a compact, single-layer attention network combined with a simple linear baseline. The linear component captures a large fraction of the variation in ionospheric parameters for short horizons (e.g., 24 hours), while the transformer provides data-adaptive corrections and multi-quantile outputs for uncertainty quantification. With a single layer in both the encoder and decoder and an embedding dimension of 128, the model contains only roughly 370K parameters. Second, the model is localized, i.e. trained on single-station inputs rather than GIMs, yet can generalize to new geographic locations. This is an important use case in edge-computing environments when the only available data may come from a local ionosonde, but accurate regional forecasting is critical. Training on single-station inputs also allows the model to leverage multiple decades of dense (15-minute cadence) ionosonde observations with lower computational overhead. Third, it is straightforward to incorporate na\"ive predictions from an existing climatological model as an additional transformer input, providing a baseline that can be adaptively corrected.

Our training procedure also differs from prior work. Many existing studies evaluate models by holding out the final one to two years of data for testing and training on earlier data. While this is common in time-series forecasting, the periodicity of the solar cycle and its influence on ionospheric dynamics mean such a test set evaluates performance only for a particular phase of solar activity. We generate training and test splits via a striping method \cite{daniell2024, smirnov2023}, using 23 years of data that span roughly two full solar cycles to produce samples throughout the entire period. The test set uses geographic locations held out entirely from training, demonstrating that the model can generalize to new locations without the need for complex graphical models.

To benchmark our forecasts, we compare LIFT against PyIRI \cite{forsythe_2024}, a Python implementation of the IRI model optimized for computational efficiency. As described in \citeA{forsythe_2024}, earlier versions of PyIRI (e.g., v0.0.4) incorporated a mixture of traditional IRI and NeQuick components using NeQuick-derived formulations for parameters such as bottomside and topside thickness, topside shape factors, and other profile characteristics. In this study, we use PyIRI v0.1.5, which has been updated to closely follow the standard IRI formulation and no longer relies on NeQuick elements. For the hmF2 parameter, PyIRI provides several options, including SHU2015, AMTB2013, and BSE-1979 \cite{Bilitza1979}; we use the SHU2015 option in this work. A detailed comparison of these submodels is available in \citeA{bilitza_2022}. PyIRI’s computational speed enables the large-scale simulations required to generate benchmark forecasts across our training and test datasets. As the most widely used empirical ionospheric model, IRI serves as a trusted standard against which new forecasting methods are commonly evaluated.

The following sections present the data used to train the model (Section \ref{sec-data}), the model architecture (Section \ref{sec-methods}), the forecasting results (Section \ref{sec-results}), and conclusions with directions for future work (Section \ref{sec-conclusions}). The results indicate that transformers may become a central component of next-generation data-driven ionospheric models. Although they lack the interpretability of full-physics systems, their ability to convert new data into accurate forecasts makes them a powerful tool for operational applications.

\section{Data}\label{sec-data}
\begin{figure}[htbp]
  \centering
  \includegraphics[width=0.95\textwidth]{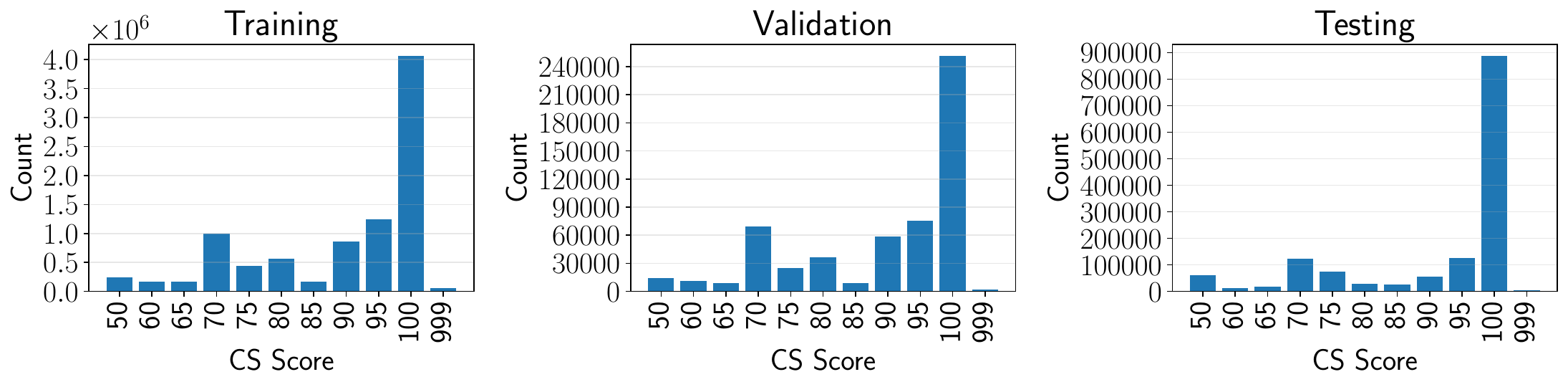}
  \caption[Distribution of ARTIST autoscaling confidence scores]{Distributions of the ARTIST confidence scores across the training, validation, and test sets.}
  \label{fig:csscores}
\end{figure}

Our model is trained using data from several openly available sources. The observational data for the target parameters, foF2, hmF2, and TEC are obtained from the Lowell GIRO Data Center digital ionogram database (DIDBase) \cite{reinisch_2011}. These parameters are determined from raw sounder data by way of an autoscaling and inversion algorithm, ARTIST5 \cite{galkin_2008}. The sounder data acquisition and subsequent parameter computation have a number of potential sources of error, which ARTIST5 attempts to quantify using a confidence score; we filter out any poorly inverted parameters using a heuristically chosen threshold, retaining only values with confidence $\ge 50$ and excluding 55. Higher confidence scores do not necessarily correspond to better autoscalings for all parameters. However, this threshold has been assessed to cover autoscalings that are accurate enough for the foF2 and hmF2 parameters; see \cite{themens_2022} for a more detailed analysis of the confidence scores of ARTIST. Figure \ref{fig:csscores} shows the distribution of confidence scores over the entire training, validation, and test sets. A future effort to train a model such as LIFT on pristine F2 and TEC parameters would be of great interest. Unfortunately, this quality of data is generally only available via manually scaled ionograms, which are not readily available in the open literature and is evident in Figure \ref{fig:csscores} by the relatively few number of code 999 observations indicating manually scaled profiles.

\begin{figure}[htbp]
  \centering
  \includegraphics[width=0.75\textwidth]{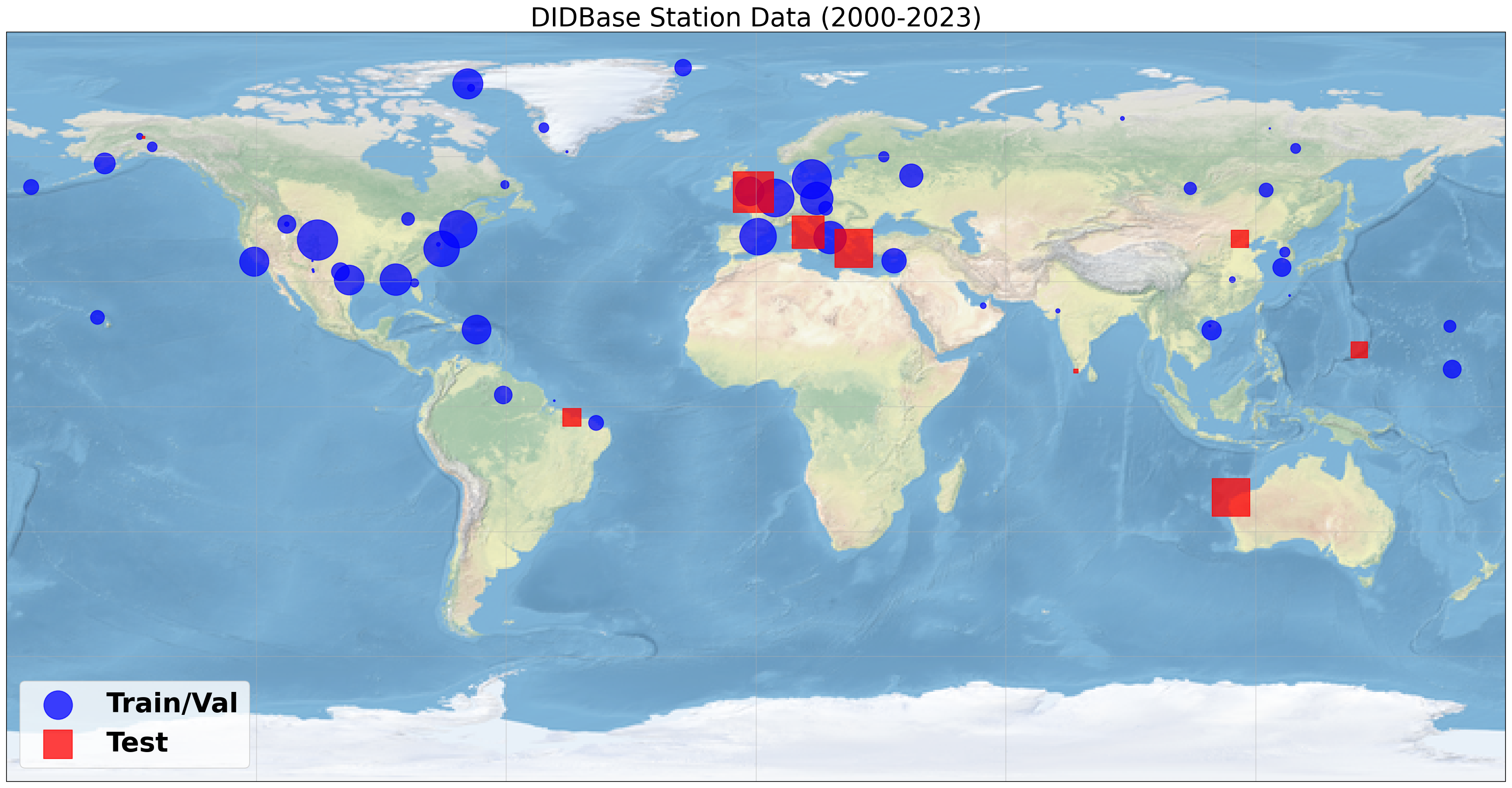}
  \caption[Geographic map of GIRO sounders]{Geographic map of GIRO stations used. The size of each marker indicates roughly the relative amount of data obtained from each location. Larger marker sizes represent more valid data obtained between the years 2000 to 2023. Some stations only had a few segments of usable data and thus are not visible in this figure.}
  \label{fig:giro_map}
\end{figure}

\begin{figure}[htbp]
  \centering
  \includegraphics[width=0.95\textwidth]{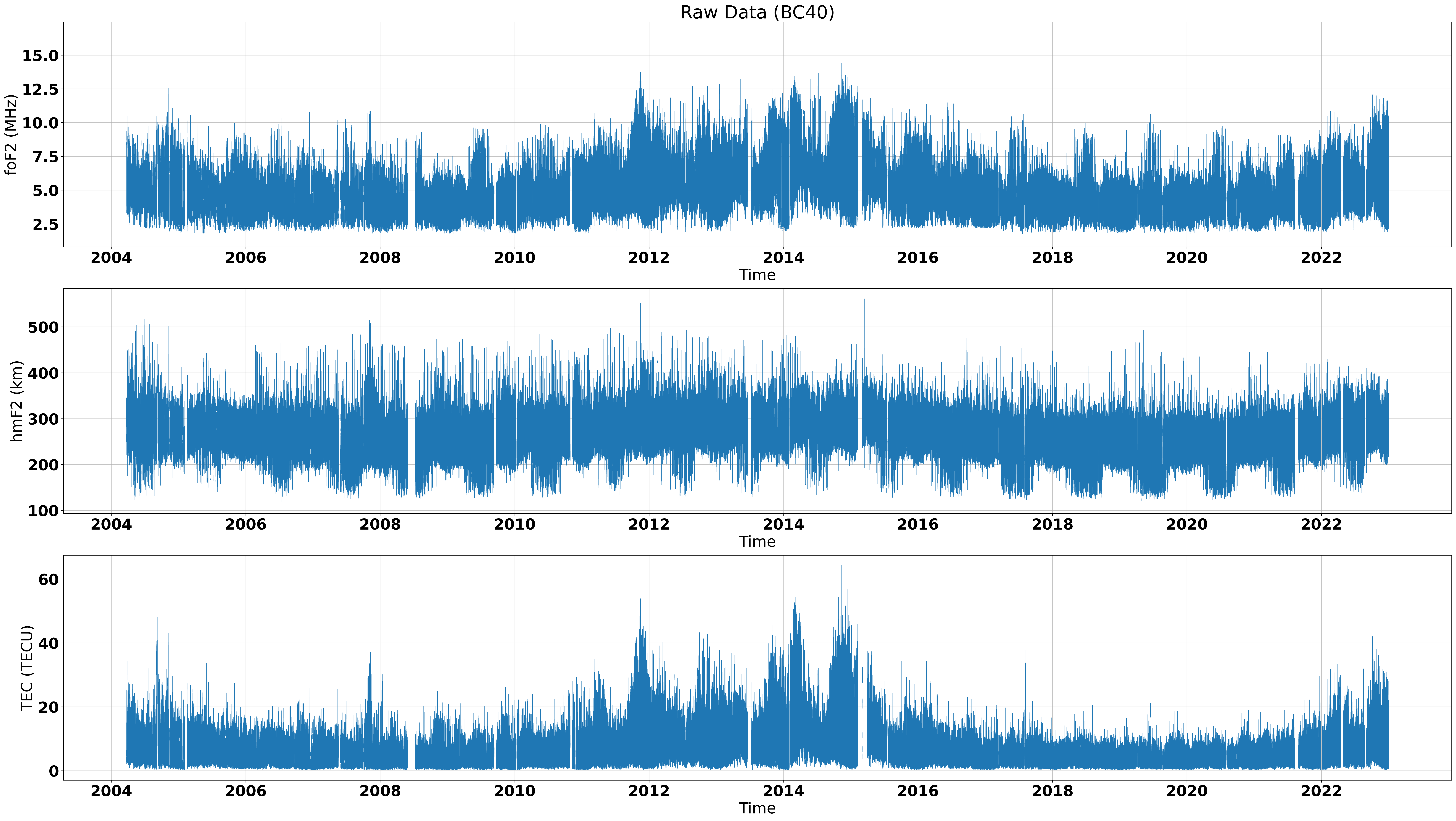}
  \caption[foF2, hmF2, and TEC parameter time series]{All data for the parameters foF2, hmF2, and TEC obtained from a Digisonde sounder in Boulder, Colorado, USA from 1 January 2000 to 1 January 2023.}
  \label{fig:raw_params}
\end{figure}

Although DIDBase provides access to data from 128 ionosonde stations, many have only sparse (infrequent) observations or poorly autoscaled data. Figure \ref{fig:giro_map} shows the geographic locations from which data were obtained, where the relative sizes of the markers indicate the amount of data from each station. Data were collected from each GIRO station for the period 1 Jan 2000 to 1 Jan 2023. The complete time series for foF2, hmF2, and TEC for one of these stations (located in Boulder, Colorado, USA) is shown in Figure \ref{fig:raw_params}. Although Boulder has comparatively dense observations, many stations have extreme gaps or very small amounts of data for much of this time period. So, in Figure \ref{fig:giro_map}, many stations simply do not show up as their relative density of data is too low.
 
Often, forecast models are tested by evaluating them on a held-out data set consisting of measurements from future time steps that the model has not seen. Unfortunately, for a localized ionospheric forecast, this tells us little about how the model will perform on data measured at a new location. Instead, we test on data from the same time periods as the training data but from completely new geographic regions; i.e., we hold out entire stations' data for testing. Figure \ref{fig:giro_map} indicates which station locations were used for training and validation versus test evaluation in our experiments. Although most available sounding measurements are located in the Northern Hemisphere, a few test stations are located near the equatorial region and the Southern Hemisphere; we reserve these held-out stations entirely for testing. This experimental design allows us to evaluate the model's ability to generalize to novel geographic locations after training. As quantified in Table \ref{tab:lowmidhigh-summary}, the LIFT model produces skillful forecasts across low, mid, and high geomagnetic latitude bands, though performance varies with ionospheric regime. This approach eliminates the need for retraining or tuning when forecasting at a new station, making the LIFT model practical for operational deployment at locations not represented in the training data.
Because we are training a local forecaster, all locations are kept in geographic latitude and longitude, and no conversions to magnetic coordinates were needed.

Our model ingests windows of data that are 72 hours in length and generates a forecast for the subsequent 24 hours; these must be contiguous periods of observations, since we wish to learn sequential temporal relationships between past measurements and covariates and future observations. After removing data with poor ARTIST5 confidence autoscalings, we further downselect to only contiguous time series from each station that contain more than the minimum number of points needed by the model, i.e., 96 hours. During this downselection process, a 96-hour window is moved across each variable of the entire dataset. If a 96-hour segment has only a few observations missing ($<10$), then we use 1-D piecewise polynomial interpolation to fill in those gaps. We ensure these missing data points do not all lie on the endpoints. Any 96-hour segment that has $>10$ missing points in either the foF2, hmF2, or TEC parameter from the sounder is not considered, and the window is moved forward.

\begin{figure}[htbp]
  \centering
  \includegraphics[width=0.75\textwidth]{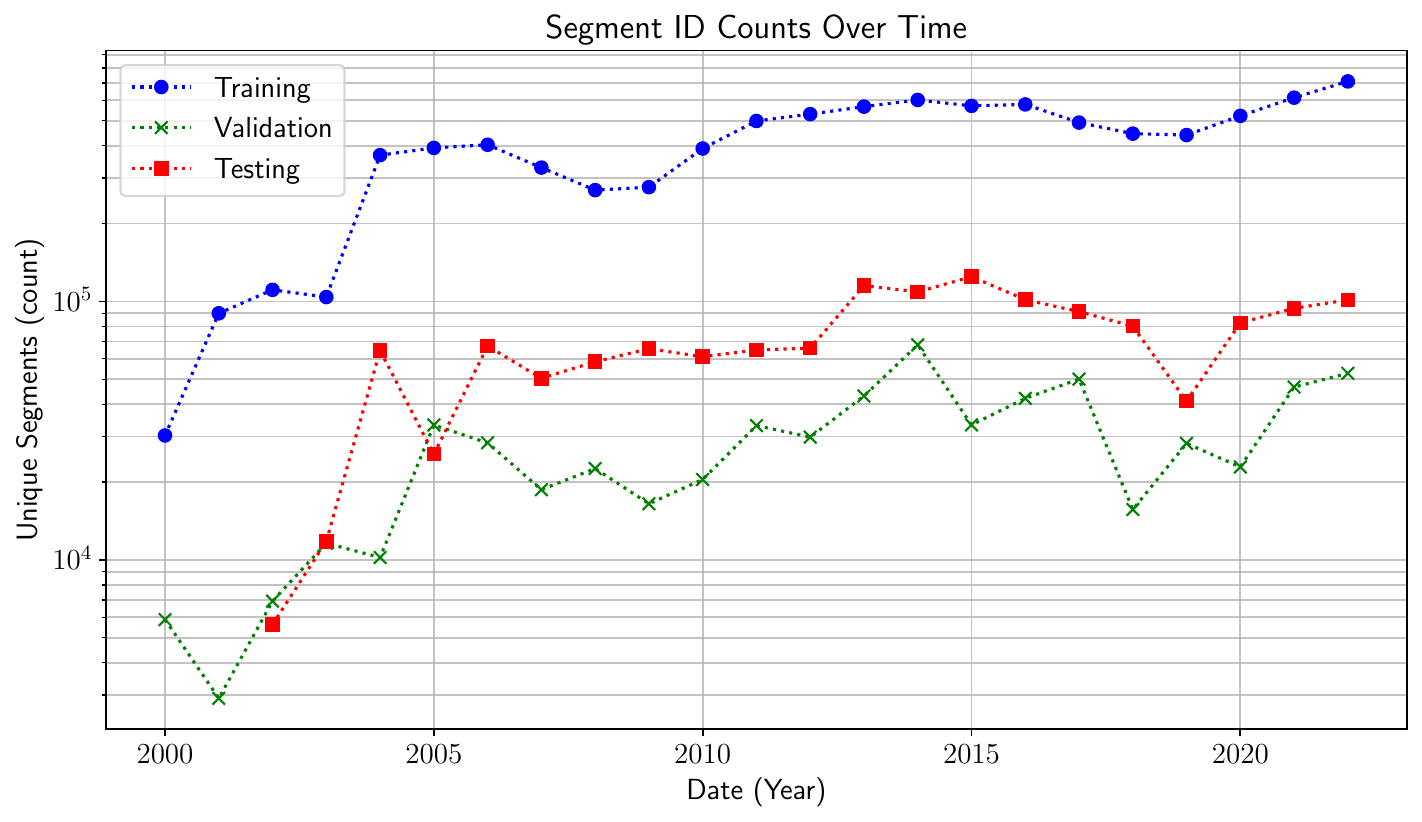}
  \caption[Scatterplot showing geographic data density]{Number of observations obtained for training (blue), validation (green), and testing (red) grouped by year.}
  \label{fig:data_years}
\end{figure}

Each of these contiguous time series is given a unique segment identifier, and this process is performed separately for the training, validation, and test sets. Figure \ref{fig:data_years} provides a summary of the number of unique sounder observations over time for the training, validation, and test sets. The total number of points and contiguous segments are reported in Table \ref{tab:num_points}.
\begin{table}[h]
    \centering
    \resizebox{0.75\textwidth}{!}{%
    \begin{tabular}{ccc}
        \hline
        & \textbf{Number of points} & \textbf{Number of 96-hour segments} \\ \hline
        \textbf{Training} & 9,383,186 & 530,677 \\ \hline
        \textbf{Validation} & 585,725 & 33,697 \\ \hline
        \textbf{Testing} & 1,494,598 & 76,894 \\ \hline
\end{tabular}
}
\caption[Summary of number of points used to train LIFT]{Total number of data points and 96-hour segments used for training, validation, and testing.}
\label{tab:num_points}
\end{table}

\begin{figure}[htbp]
  \centering
  \includegraphics[width=0.95\textwidth]{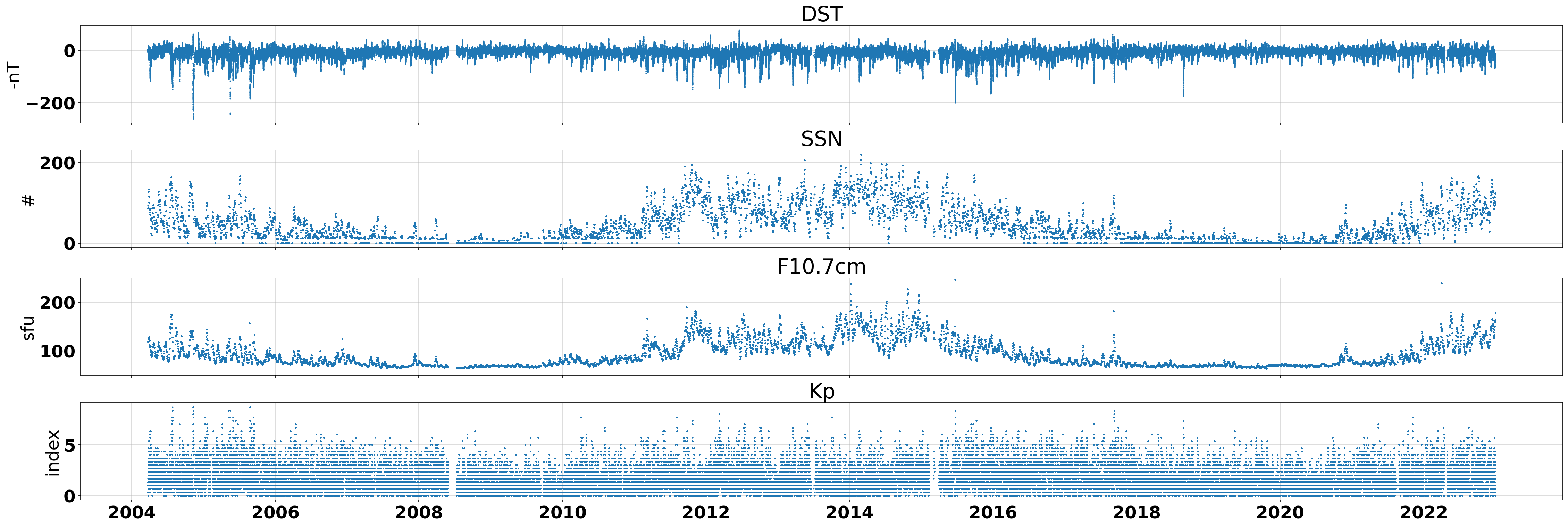}
  \caption[Exogenous inputs for LIFT]{Indices used as exogenous inputs to the model for the time period 2004-2023.}
  \label{fig:data_indices}
\end{figure}

Each data segment is then paired with a corresponding set of drivers that are used as exogenous inputs to the model. These drivers include sunspot number (SSN), the Kp index (Kp), and F10.7 cm solar flux, which were obtained from the GFZ German Research Center for Geosciences \cite{matzka_2021}, and the Dst index from the WDC for Geomagnetism, Kyoto \cite{nose_2015}. Together, these indices represent a set of additional covariates that can be measured or obtained alongside new sounding data, but whose interactions with the ionospheric parameters may be too complex, nonlinear, or lagged to be expressed as closed analytic expressions. Figure \ref{fig:data_indices} illustrates the affiliated time series of these indices for the Boulder, CO, USA ionosonde station.

All input variables were aligned to a common 15-minute temporal cadence. Ionosonde observations of foF2, hmF2, and TEC were obtained at this native sampling rate from the GIRO repository. Geomagnetic and solar indices that are reported at coarser temporal resolutions were forward-filled to match this cadence: Dst values (1-hour resolution) were repeated across four consecutive time steps, Kp values (3-hour resolution) across twelve time steps, and SSN (daily resolution) across all 96 time steps in each day. This forward-filling approach is consistent with operational practice, where the most recently available index value is applied until a new measurement is reported. Covariates that can be computed analytically, including solar zenith angle and PyIRI climatological predictions, were evaluated directly at each 15-minute time step. The transformer's self-attention mechanism learns to weight inputs according to their relevance to the forecast, so this temporal alignment does not require uniform information content across all time steps; the model can learn to attend preferentially to time steps at which coarse-cadence indices are updated.

\section{Methods}\label{sec-methods}
The complexities and multiscale nature of the ionospheric data require an architecture that is flexible enough to deal with the varying space weather conditions across different latitudes and seasons. In this section, we introduce our transformer-based model, which is designed to leverage these complex datasets for probabilistic forecasting. First, we establish some basic notation and expressions, denoting a time series of a single variable from time index $t$ to $t+k$ as a vector $\mathbf{x}_{t:t+k} \in \mathbb{R}^{k+1}$, and a time series of $m$ variables as a matrix $\mathbf{X}_{t:t+k} \in \mathbb{R}^{(k+1) \times m}$. We will refer to scalars, such as the $j^{th}$ variable at time $t+k$ as $x_{t+k;j}$.

We then define the data used in our model as follows:
\begin{align}
    &\mathbf{X}_{t-c:t} \in \mathbb{R}^{(c+1) \times d_1}: &&\text{Past observations of the target variables}\\
    &\mathbf{Y}_{t-c:t} \in \mathbb{R}^{(c+1) \times d_2}: &&\text{Past observations of the covariates}\\
    &\mathbf{Z}_{t-c:t+p} \in \mathbb{R}^{(c+p+1) \times d_3}: &&\text{Past and future covariates computable up to $t+p$}
\end{align}
where $c$ and $p$ are the number of \emph{context} and \emph{prediction} points, respectively. The matrix $\mathbf{X}$ contains the foF2, hmF2, and TEC parameters observed from an ionosonde, thus $d_1 = 3$, $\mathbf{Y}$ corresponds to the data in Figure \ref{fig:data_indices} and consists of the exogenous or forcing variables that we expect to influence the forecast, so $d_2=4$, and $\mathbf{Z}$ contains data for computable covariates, including a time vector and the na\"ive PyIRI prediction, which we describe in more detail in Sections \ref{sec-pos_encoding} and \ref{sec-model_covariates}, resulting in $d_3=13$.

We seek a model for the conditional distribution of each target variable at each point in the forecast horizon from which we may estimate their expected values and provide uncertainty quantification. A common method is to assume a fixed parametric form for the distributions, e.g.
\begin{equation}\label{eqn:pdf}
    f_{t+k;j}(x_{t+k;j} \mid \mathbf{X}_{t-c:t}, \mathbf{Y}_{t-c:t}, \mathbf{Z}_{t-c:t+p}) \sim \mathcal{N}(\mu_{t+k;j}, \sigma_{t+k;j}),
\end{equation}
where $j\in \{1,...,d_1\}$ indicates the target variable, and $\mu_{t+k;j}$ and $\sigma_{t+k;j}$ are predicted parameters of a normal distribution $k$ points into the future. 
However, such a representation is sensitive to the selected parametric form and may perform poorly due to model mismatch.
Instead, we elect to implement a fully nonparametric approach using quantiles. To this end, let the conditional cumulative distribution function (CDF) of Equation \ref{eqn:pdf} be
\begin{equation}
    F_{t+k;j}(x_{t+k;j} \mid \mathbf{X}_{t-c:t}, \mathbf{Y}_{t-c:t}, \mathbf{Z}_{t-c:t+p}) \coloneqq \text{Pr}(x_{t+k;j} \le x \mid \mathbf{X}_{t-c:t}, \mathbf{Y}_{t-c:t}, \mathbf{Z}_{t-c:t+p}) = \tau
\end{equation}
where $\tau \in [0, 1]$. Then, the \emph{quantile function} is the inverse of our conditional CDF
\begin{align}\label{eqn:quantilefunction}
\begin{split}
    Q_{x_{t+k;j} \mid \mathbf{X}_{t-c:t}, \mathbf{Y}_{t-c:t}, \mathbf{Z}_{t-c:t+p}}(\tau) &= F_{t+k;j}^{-1}\left(x_{t+k;j} \mid \mathbf{X}_{t-c:t}, \mathbf{Y}_{t-c:t}, \mathbf{Z}_{t-c:t+p}\right) \\
     &= \inf \left\{ x \in \mathbb{R} : F_{t+k;j}(x \mid \mathbf{X}_{t-c:t}, \mathbf{Y}_{t-c:t}, \mathbf{Z}_{t-c:t+p}) \geq \tau \right\} \\
     &= x_{t+k;j}^{\tau}.
\end{split}
\end{align} 
The infimum in this definition is necessary only if the CDF happens to be constant on some interval, in which case $Q_{x_{t+k;j}\mid \mathbf{X}_{t-c:t}, \mathbf{Y}_{t-c:t}, \mathbf{Z}_{t-c:t+p}}(\tau)$ would not be well defined \cite{casella_2002, gilchrist_2000, gneiting_2016}. 

To estimate these functions, we construct a model with two main components: a linear predictor, which is designed to capture linear trends and the diurnal pattern from observations of the target variable, and a transformer neural network, which learns nonlinear relationships across all past and future data. The output of the transformer is a set of values that together with the linear component allows us to predict $Q_{\cdot \mid \cdot}(\tau)$ for many different values of $\tau$. So, we have,
\begin{equation}\label{eqn:themodel}
    \hat{\mathbf{x}}_{t+1:t+p;j}^{\tau} = 
    \underbrace{L\left(\mathbf{x}_{t-c:t;j}\right)}_{\text{Linear}} 
    + 
    \underbrace{T\left(\mathbf{X}_{t-c:t}, \mathbf{Y}_{t-c:t}, \mathbf{Z}_{t-c:t+p}; \tau, j \right)}_{\text{Transformer}},
\end{equation}
where $\hat{\mathbf{x}}_{t+1:t+p;j}^{\tau}$ is the forecasted time series of either foF2, hmF2, or TEC, depending on $j$ for the specified quantile $\tau$ over the prediction horizon $t+1$ to $t+p$. The function $L(\cdot)$ is straightforward and is defined in the remainder of this section, while $T(\cdot)$ will be defined in more detail in Section \ref{sec-transformercomp}. The loss function used to train this model will be provided in Section \ref{sec-loss_fun}.

\begin{figure}[htbp]
  \centering
  \includegraphics[width=1.0\textwidth]{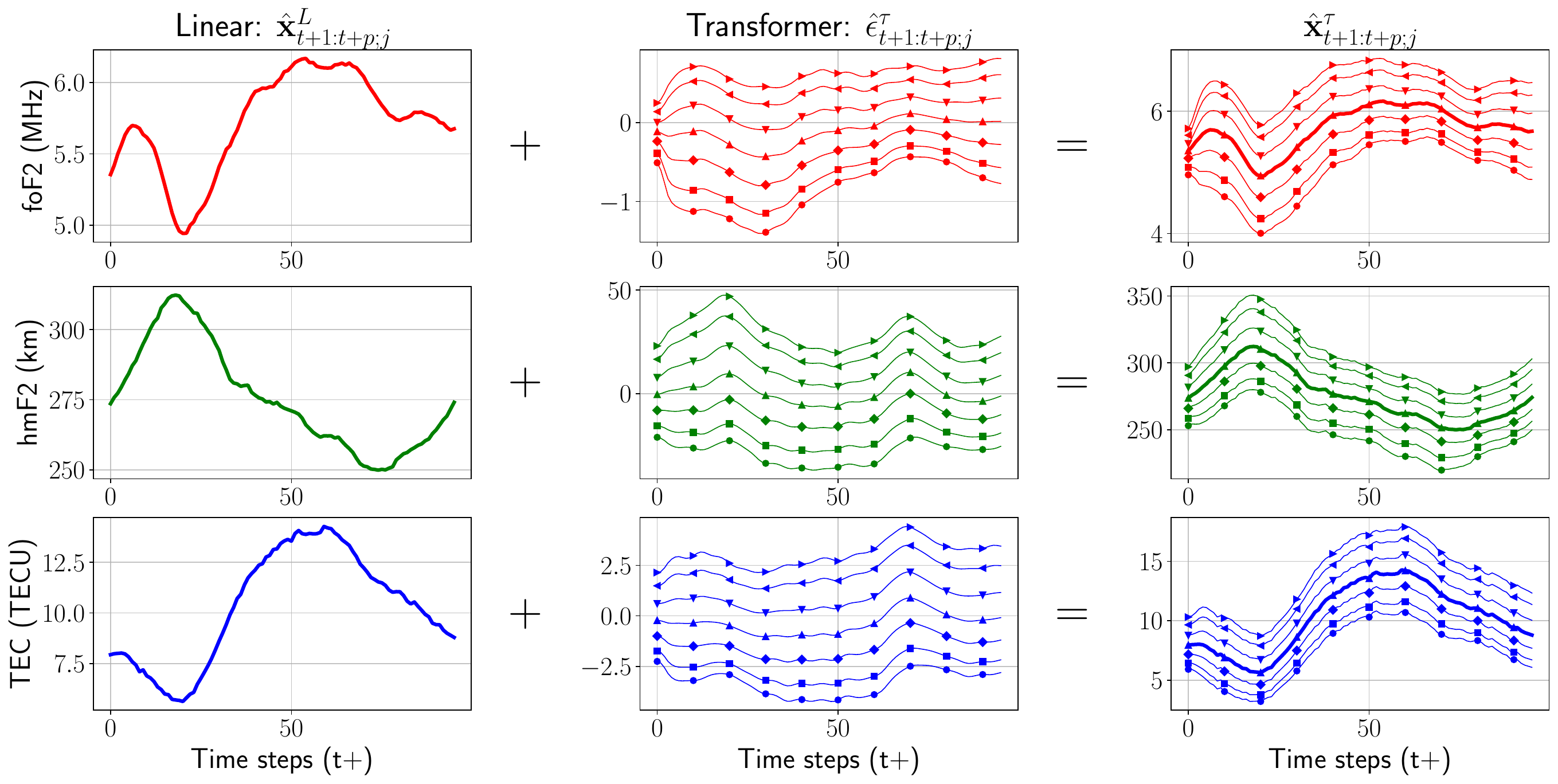}
  \caption[LIFT model components]{An illustration of how the LIFT model generates point forecasts and uncertainty quantification. Each row provides results for a different target variable (foF2, hmF2, TEC). The linear (first column) and nonlinear (second column) components add together to generate an estimate of the quantile distribution for each target (third column). The final median prediction for each target variable is represented by the bold weighted lines in the third column.}
  \label{fig:model_components}
\end{figure}

\begin{figure}[htbp]
  \centering
  \includegraphics[width=0.85\textwidth]{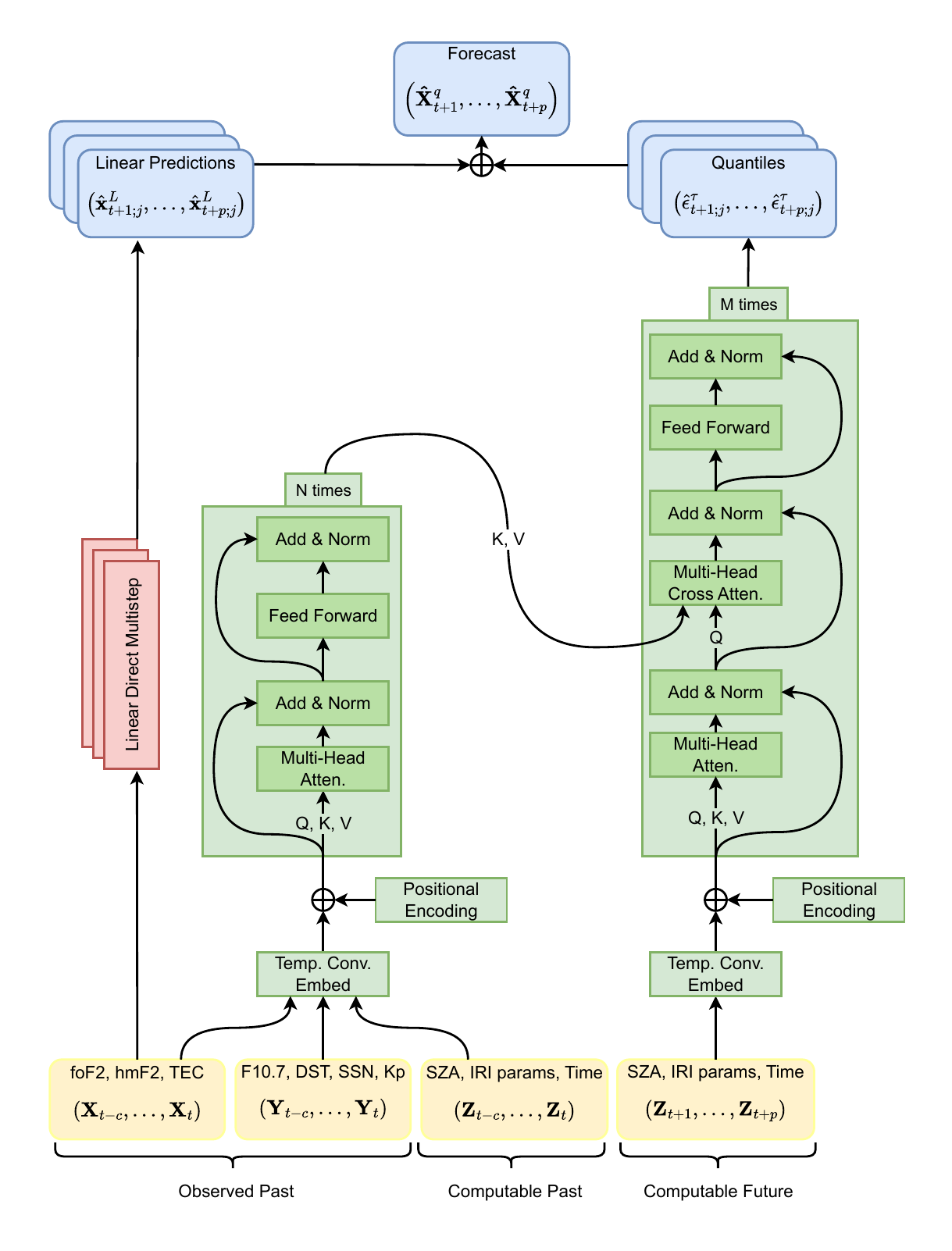}
  \caption[LIFT model diagram]{Complete LIFT model diagram. All components of the model related to the transformer are shaded green, while the linear model is indicated by red, the inputs in yellow, and the outputs in blue. The stacks each correspond to an individual forecasted parameter (foF2, hmF2, and TEC), so there are three sets of outputs from the linear models and the transformer. The final forecast data at the top is a matrix containing all aggregated outputs. We include the $N$ and $M$ blocks for the transformer blocks for completeness. The LIFT model uses $N=M=1$.}
  \label{fig:model_diagram}
\end{figure}

The linear component is used to predict the simple linear trends that we expect to see in each of the target parameters by generating predictions for the next $p$ time steps based solely on the previous $c$ time steps. Furthermore, it predicts all $p$ time steps simultaneously, similar to sequence-to-sequence models \cite{sutskever_2014}, and so we refer to it as a \emph{direct multistep} linear model. We fit independent linear models for each of the $j$ target variables (foF2, hmF2, TEC),
\begin{align}\label{eqn:linearmodel}
\begin{split}
    &\hat{\mathbf{x}}_{t+1:t+p;j}^{L} =
    \mathbf{W}_{j}\mathbf{x}_{t-c:t;j} + \mathbf{b}_{j},
\end{split}
\end{align}
with the superscript $L$ denoting this prediction comes from the linear model. The weight matrices $\mathbf{W}_{j} \in \mathbb{R}^{p \times (c+1)}$ and bias vectors $\mathbf{b}_{j} \in \mathbb{R}^{p}$ are fit using the Adam optimizer \cite{kingma_2014} alongside the transformer network. So, this model only generates a linear approximation for a vector of future values from a vector of past values. This forecast will have residual errors,
\begin{equation}\label{eqn:linerrors}
    \boldsymbol{\epsilon}_{t+1:t+p;j} = \mathbf{x}_{t+1:t+p;j} - \hat{\mathbf{x}}_{t+1:t+p;j}^{L},
\end{equation}
each of which follow some unknown distribution that we may reasonably expect to depend on a set of drivers of the system. Instead of approximating the distributions for these errors, we apply a transformer model to learn quantiles for these residuals directly,
\begin{equation}\label{eqn:residualquantiles}
    \hat{\boldsymbol{\epsilon}}_{t+1:t+p;j}^{\tau} = T\left(\mathbf{X}_{t-c:t},\mathbf{Y}_{t-c:t},\mathbf{Z}_{t-c:t+p}; \tau, j \right).
\end{equation}
In our model, we fit seven different quantile levels,  $\tau \in \{0.05, 0.1, 0.25, 0.5, 0.75, 0.9, 0.95\}$. Then, a point forecast from our model is obtained by adding the linear component to the predicted median, $\tau=0.5$, from this transformer,
\begin{align}\label{eqn:medpred}
    \hat{\mathbf{x}}_{t+1:t+p;j}^{0.5} = \hat{\mathbf{x}}_{t+1:t+p;j}^{L} + \hat{\boldsymbol{\epsilon}}_{t+1:t+p;j}^{0.5} = L\left(\mathbf{x}_{t-c:t;j}\right) + T\left(\mathbf{X}_{t-c:t}, \mathbf{Y}_{t-c:t}, \mathbf{Z}_{t-c:t+p}; 0.5, j\right).
\end{align}
Moreover, we are able to construct confidence intervals (CI) using different levels of $\tau$. For example, the upper and lower bounds of a $90\%$-CI are generated using
\begin{align}\label{eqn:cipred}
    \hat{\mathbf{x}}_{t+1:t+p;j}^{\text{0.95}} &= \hat{\mathbf{x}}_{t+1:t+p;j}^{L} + \hat{\boldsymbol{\epsilon}}_{t+1:t+p;j}^{0.95} = L\left(\mathbf{x}_{t-c:t;j}\right) + T\left(\mathbf{X}_{t-c:t}, \mathbf{Y}_{t-c:t}, \mathbf{Z}_{t-c:t+p}; 0.95, j\right)\\
    \hat{\mathbf{x}}_{t+1:t+p;j}^{\text{0.05}} &= \hat{\mathbf{x}}_{t+1:t+p;j}^{L} + \hat{\boldsymbol{\epsilon}}_{t+1:t+p;j}^{0.05} = L\left(\mathbf{x}_{t-c:t;j}\right) + T\left(\mathbf{X}_{t-c:t}, \mathbf{Y}_{t-c:t}, \mathbf{Z}_{t-c:t+p}; 0.05, j\right).
\end{align}
Conceptually, the transformer acts as a nonlinear correction to the simple linear model that incorporates all available data, including drivers and na\"ive IRI-based climatology, but without the need to specify explicit functional relationships to the target parameters foF2, hmF2, and TEC. Figure \ref{fig:model_components} provides a concrete example of the output of each of the LIFT model components, and Figure \ref{fig:model_diagram} illustrates the complete end-to-end architecture of our LIFT model architecture.

\subsection{Transformer Component}\label{sec-transformercomp}
Our transformer is based on the original encoder-decoder architecture introduced by \cite{vaswani_2017} which uses multi-head self-attention networks followed by fully connected feed-forward layers. The attention mechanism allows the transformer to discover long-range temporal relationships very efficiently and to incorporate information from the covariate time series into its forecast of each quantile in Equation \ref{eqn:residualquantiles}. This is a desirable property for ionospheric forecasting, since there may be unknown time lags between changes in global driver indices and their effects on the target parameters \cite{schmolter_2024, chen_2018, chen_2015, coley_2012, jakowski_1991}.

In our adaptation for time series forecasting, we introduce 1-dimensional convolutions that embed each set of variables in a $d_m$-dimensional latent space, allowing the model to learn richer, high-dimensional feature representations for both the encoder and the decoder inputs. For our model, we use $d_m=128$, and a kernel size of 1 time step, so the embedding is simply mapping the data from each time step into latent coordinates,
\begin{align}
\begin{split}
    &E_{\text{enc}}: \mathbb{R}^{d_1+d_2+d_3} \rightarrow \mathbb{R}^{d_m},\\
    &E_{\text{dec}}: \mathbb{R}^{d_3} \rightarrow \mathbb{R}^{d_m},
\end{split}
\end{align}
or, when applied to our time series,
\begin{align}
    &E_{\text{enc}}(\mathbf{X}_{t-c:t}, \mathbf{Y}_{t-c:t}, \mathbf{Z}_{t-c:t}) = \tilde{\mathbf{X}}_{t-c:t} \in \mathbb{R}^{(c+1) \times d_m}, \label{eqn:encembed}\\
    &E_{\text{dec}}(\mathbf{Z}_{t+1:t+p}) = \tilde{\mathbf{X}}_{t+1:t+p} \in \mathbb{R}^{p \times d_m}.\label{eqn:decembed}
\end{align}

The transformer is composed of three layers of attention which we will call the encoder self-attention, decoder self-attention, and decoder cross-attention. Each layer consists of $H$ \emph{attention heads}. We will define the process for computing the encoder and decoder self-attention first, since the cross attention uses output from the previous two. These first two layers begin with a set of learned projections,
\begin{align}
    &\mathbf{W}_{Q,h,enc}, \mathbf{W}_{K,h,enc}, \mathbf{W}_{V,h,enc} \in \mathbb{R}^{(d_m \times d_k)},\label{eqn:encprojections}\\
    &\mathbf{W}_{Q,h,dec}, \mathbf{W}_{K,h,dec}, \mathbf{W}_{V,h,dec}\in \mathbb{R}^{(d_m \times d_k)},\label{eqn:decprojections}
\end{align}
where $d_k=d_m/H$ is the latent dimension for each head and $h$ is the head index. These project each latent vector onto what are known as \emph{query}, \emph{key}, and \emph{value} vectors. We represent time series of these query, key, and value vectors as matrices by stacking the vectors row-wise. So, when the projections in Equations \ref{eqn:encprojections} and \ref{eqn:decprojections} are applied to our embedded time series in Equations \ref{eqn:encembed} and \ref{eqn:decembed}, respectively, we have
\begin{align}
\begin{split}
    &\mathbf{Q}_{h,\text{enc}} = \tilde{\mathbf{X}}_{t-c:t} \mathbf{W}_{Q,h,\text{enc}}, \\
    &\mathbf{K}_{h,\text{enc}} = \tilde{\mathbf{X}}_{t-c:t} \mathbf{W}_{K,h,\text{enc}}, \\
    &\mathbf{V}_{h,\text{enc}} = \tilde{\mathbf{X}}_{t-c:t} \mathbf{W}_{V,h,\text{enc}},
\end{split}
\end{align}
where $\mathbf{Q}_{h,\text{enc}}, \mathbf{K}_{h,\text{enc}}, \mathbf{V}_{h,\text{enc}} \in \mathbb{R}^{(c+1) \times d_k}$, and
\begin{align}
\begin{split}
    &\mathbf{Q}_{h,\text{dec}} = \tilde{\mathbf{X}}_{t+1:t+p} \mathbf{W}_{Q,h,\text{dec}}, \\
    &\mathbf{K}_{h,\text{dec}} = \tilde{\mathbf{X}}_{t+1:t+p} \mathbf{W}_{K,h,\text{dec}}, \\
    &\mathbf{V}_{h,\text{dec}} = \tilde{\mathbf{X}}_{t+1:t+p} \mathbf{W}_{V,h,\text{dec}},
\end{split}
\end{align}
where $\mathbf{Q}_{h,\text{dec}}, \mathbf{K}_{h,\text{dec}}, \mathbf{V}_{h,\text{dec}} \in \mathbb{R}^{p \times d_k}$. The queries and keys are then used to compute the attention scores,
\begin{align}\label{eqn:scores}
    &\mathbf{S}_{h,\text{enc}} = \frac{\mathbf{Q}_{h, \text{enc}}\mathbf{K}_{h,\text{enc}}^{T}}{\sqrt{d_k}} \in \mathbb{R}^{(c+1) \times (c+1)}\\
    &\mathbf{S}_{h,\text{dec}} = \frac{\mathbf{Q}_{h, \text{dec}}\mathbf{K}_{h,\text{dec}}^{T}}{\sqrt{d_k}} \in \mathbb{R}^{p \times p}.
\end{align}
In principle, these scores determine how much weight, or attention, the model should place on each time step of its input data when producing each time step of its output, and the division by $\sqrt{d_k}$ simply prevents the scores from growing too large if the embedding dimension is increased. 

We then pass these scores through Softmax functions which convert each score vector into probabilities and multiply by the affiliated value matrices,
\begin{align}\label{eqn:attentions}
    &\mathbf{A}_{h,\text{enc}} = \text{Softmax}(\mathbf{S}_{h,\text{enc}})\mathbf{V}_{h,\text{enc}} \in \mathbb{R}^{(c+1)\times d_k}, \\
    &\mathbf{A}_{h,\text{dec}} = \text{Softmax}(\mathbf{S}_{h,\text{dec}})\mathbf{V}_{h,\text{dec}} \in \mathbb{R}^{p \times d_k}.
\end{align}
Each of these matrices, $\mathbf{A}_{\cdot, \cdot}$, use full self-attention, which means that the attention score for any time step can incorporate values from anywhere in its input time series. Each set of attention heads is then concatenated and projected back to the original embedding dimension,
\begin{align}
    &\bar{\mathbf{X}}_{t-c:t} = [\mathbf{A}_{1,\text{enc}} \mathbf{A}_{2,\text{enc}} \dots \mathbf{A}_{H,\text{enc}}] \mathbf{W}_{O,\text{enc}}, &&\mathbf{W}_{O,\text{enc}}\in \mathbb{R}^{Hd_k \times d_m} \label{eqn:transconcat1}\\
    &\bar{\mathbf{X}}_{t+1:t+p} = [\mathbf{A}_{1,\text{dec}} \mathbf{A}_{2,\text{dec}} \dots \mathbf{A}_{H,\text{dec}}] \mathbf{W}_{O,\text{dec}}, &&\mathbf{W}_{O,\text{dec}}\in \mathbb{R}^{Hd_k \times d_m},\label{eqn:transconcat2}
\end{align}
which allows the model to share information across the heads and produce an output that is a time series in the $d_m-$dimensional model space. 

After each attention layer, we apply a fully connected feed-forward network with a nonlinear activation. For our model, we use GELU activation functions \cite{hendrycks_2023}. This choice of activation function is simply because GELU provides modest training stability and convergence in other transformer applications; however, a ReLU function would likely work as well here. Although the original transformer architecture used different hidden dimensions in the feed-forward layer, $d_{ff}$, than in the attention mechanisms, we found that our model performed well with $d_{ff}=d_{m}=128$. These feed-forward networks introduce additional nonlinearity to the transformer's processing of the data through their activation functions,
\begin{align}
    &\bar{\mathbf{X}}'_{t-c:t} = \sigma(\bar{\mathbf{X}}_{t-c:t}\mathbf{W}_{ff1,\text{enc}} + \mathbf{b}_{ff1,\text{enc}})\mathbf{W}_{ff2,\text{enc}} + \mathbf{b}_{ff2, \text{enc}}, \label{eqn:encout}\\
    &\bar{\mathbf{X}}'_{t+1:t+p} = \sigma(\bar{\mathbf{X}}_{t+1:t+p}\mathbf{W}_{ff1,\text{dec}} + \mathbf{b}_{ff1,\text{dec}})\mathbf{W}_{ff2,\text{dec}} + \mathbf{b}_{ff2, \text{dec}},\label{eqn:decout}
\end{align}
where $\sigma(\cdot)$ is the nonlinear activation. We also incorporate normalization layers \cite{ba_2016} after each attention and feed-forward step to improve the training process. 

The output from Equations \ref{eqn:encout} and \ref{eqn:decout} are now ready for processing in the decoder's cross-attention. Here, a new set of queries, keys, and values are created. The queries use output from the decoder's self-attention, and the keys and values use output from the encoder's self-attention,
\begin{align}\label{eqn:crossprojections}
\begin{split}
    &\mathbf{Q}_{h,\text{cross}} = \bar{\mathbf{X}}^{'}_{t+1:t+p} \mathbf{W}_{Q,h,\text{cross}},\\
    &\mathbf{K}_{h,\text{cross}} = \bar{\mathbf{X}}^{'}_{t-c:t} \mathbf{W}_{K,h,\text{cross}},\\
    &\mathbf{V}_{h,\text{cross}} = \bar{\mathbf{X}}^{'}_{t-c:t} \mathbf{W}_{V,h,\text{cross}},
\end{split}
\end{align}
where $\mathbf{Q}_{h,\text{cross}} \in \mathbb{R}^{p \times d_k}$ and $\mathbf{K}_{h,\text{cross}}, \mathbf{V}_{h,\text{cross}} \in \mathbb{R}^{(c+1) \times d_k}$. The cross-attention scores are given by
\begin{align}
    &\mathbf{S}_{h,cross} = \frac{\mathbf{Q}_{h,\text{cross}}\mathbf{K}^{T}_{h,\text{cross}}}{\sqrt{d_k}} \in \mathbb{R}^{p \times (c+1)}\\
    &\mathbf{A}_{h,\text{cross}} = \text{Softmax}\left(\mathbf{S}_{h,cross}\right)\mathbf{V}_{h,\text{cross}} \in \mathbb{R}^{p \times d_k},
\end{align}
and the concatenated heads are projected back into the model latent dimension,
\begin{align}\label{eqn:transconcat3}
    &\bar{\bar{\mathbf{X}}}_{t+1:t+p} = [\mathbf{A}_{1,\text{cross}} \mathbf{A}_{2,\text{cross}} \dots \mathbf{A}_{H,\text{cross}}]\mathbf{W}_{O,\text{cross}}, &&\mathbf{W}_{O,\text{cross}}\in \mathbb{R}^{Hd_k \times d_m},
\end{align}
where the double bar accent indicates this output is distinct from Equation \ref{eqn:transconcat2}. As with the self-attention layers, the cross-attention layer includes a final feed-forward network,
\begin{equation}
     \bar{\bar{\mathbf{X}}}^{'}_{t+1:t+p} = \sigma(\bar{\bar{\mathbf{X}}}_{t+1:t+p}\mathbf{W}_{ff1,\text{cross}} + \mathbf{b}_{ff1,\text{cross}})\mathbf{W}_{ff2,\text{cross}} + \mathbf{b}_{ff2, \text{cross}}.\label{eqn:deccrossout}
\end{equation}
The LIFT model requires one final set of maps to transform the decoder's cross-attention output into the predicted quantiles for foF2, hmF2, and TEC at each point from $t+1$ to $t+p$, 
\begin{align}
    &\hat{\mathbf{E}}_{t+1:t+p}^{\text{foF2}} = \bar{\bar{\mathbf{X}}}^{'}_{t+1:t+p}\mathbf{W}_{\text{foF2}} + \mathbf{b}_{\text{foF2}}, &&\hat{\mathbf{E}}_{t+1:t+p}^{\text{foF2}} \in \mathbb{R}^{p \times 7} \label{eqn:quantileoutfof2}\\
    &\hat{\mathbf{E}}_{t+1:t+p}^{\text{hmF2}} = \bar{\bar{\mathbf{X}}}^{'}_{t+1:t+p}\mathbf{W}_{\text{hmF2}} + \mathbf{b}_{\text{hmF2}}, &&\hat{\mathbf{E}}_{t+1:t+p}^{\text{hmF2}} \in \mathbb{R}^{p \times 7} \label{eqn:quantileouthmf2}\\
    &\hat{\mathbf{E}}_{t+1:t+p}^{\text{TEC}} = \bar{\bar{\mathbf{X}}}^{'}_{t+1:t+p}\mathbf{W}_{\text{TEC}} + \mathbf{b}_{\text{TEC}}, &&\hat{\mathbf{E}}_{t+1:t+p}^{\text{TEC}} \in \mathbb{R}^{p \times 7} \label{eqn:quantileoutTEC}.
\end{align}
Each column in $\hat{\mathbf{E}}_{t+1:t+p}^{foF2}, \hat{\mathbf{E}}_{t+1:t+p}^{hmF2}$, and $\hat{\mathbf{E}}_{t+1:t+p}^{TEC}$ forms a forecast for one of the $7$ different residual quantile levels for each parameter, respectively, i.e. we complete our model for Equation \ref{eqn:residualquantiles}.

\begin{figure}[htbp]
    \centering
    \begin{tabular}{c}
        \includegraphics[width=1\textwidth]{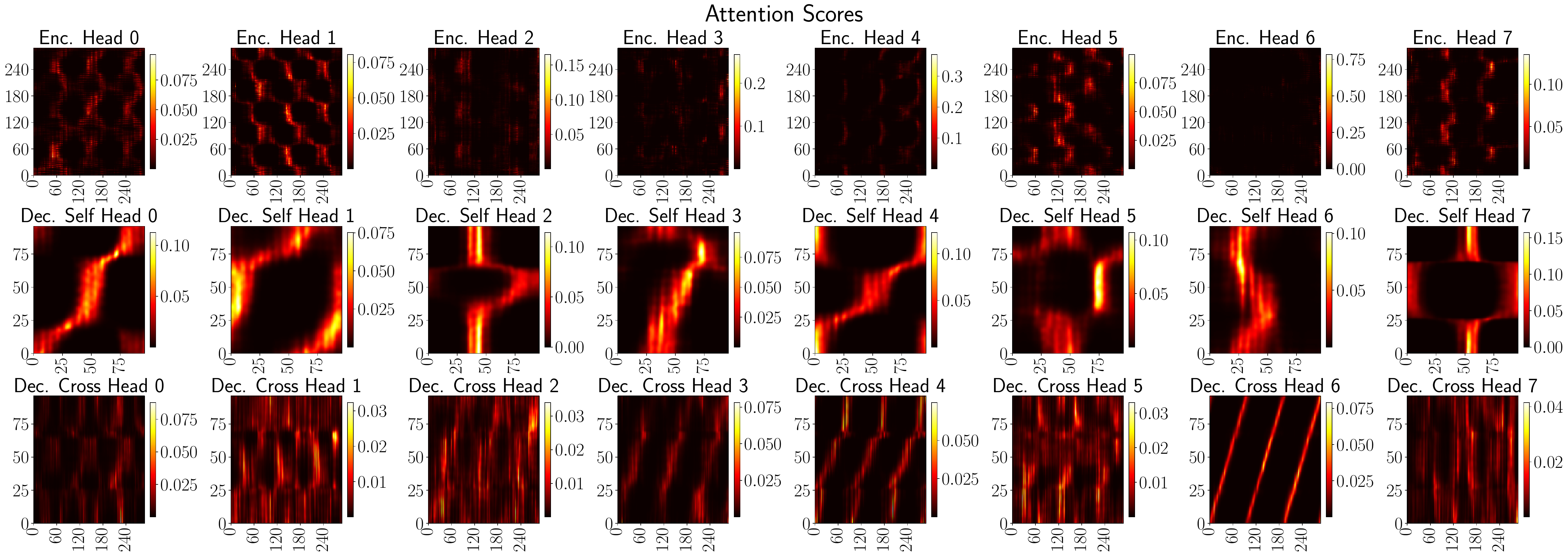}
    \end{tabular}
    \caption[Transformer attention plots]{A visualization of the attention scores for the encoder self-attention (top), decoder self-attention (middle), and decoder cross-attention (bottom) for a randomly selected test sequence.}
    \label{fig:attention_plots}
\end{figure}

The decoder's cross-attention layer performs the critical function of fusing the information from the encoder with that of the decoder. Larger transformer models may include several \emph{blocks} of encoders and decoders, each of which computes a new set of attention heads and repeats the entire process outlined above. These models will have many more parameters to train and thus require very large amounts of data; however, they offer more flexibility by allowing each new set of attention heads in each layer to attend to different information. For our model, we found that a single block of attention heads in both our encoder and decoder performed well and kept the total number of trainable parameters much lower than other transformer implementations for time series forecasting \cite{zhou_2021, lim_2021}. Figure \ref{fig:attention_plots} provides a visualization of the attention scores for each of the model's layers for an encoder input length of $c=288$ and a decoder input length of $p=96$. This corresponds to a context length of 3 days and a prediction length of 1 day.

A final note is that, in practice, transformers often utilize residual connections, as introduced by \cite{he_2016}, which simply add the input and output of a given layer together before being passed into the following layer, e.g. Equation \ref{eqn:encout} would be implemented as
\begin{equation}
    \bar{\mathbf{X}}'_{t-c:t} = \sigma((\bar{\mathbf{X}}_{t-c:t} + \tilde{\mathbf{X}}_{t-c:t})\mathbf{W}_{ff1,\text{enc}} + \mathbf{b}_{ff1,\text{enc}})\mathbf{W}_{ff2,\text{enc}} + \mathbf{b}_{ff2, \text{enc}}.
\end{equation}
This is now common practice in machine learning models that involve many layers of neural networks, as residual connections have been known to help alleviate vanishing gradients and improve overall stability during training \cite{he_2020}. However, because our model consists of a single block of attention layers, i.e. one pass through the encoder and one pass through the decoder, these skip connections are not required. Nevertheless, we include them here since future efforts in the space weather community may require more expressive models, and one way of achieving this is by including additional blocks. However, additional attention blocks will greatly increase the parameter count of the model.

\subsection{Loss Function}\label{sec-loss_fun}
The loss function uses a combination of the mean squared error (MSE) of the linear predictors and the multi-quantile loss on the combined linear and residual predictions. The quantile loss, or pinball loss, for a single quantile $\tau \in (0,1)$ is
\begin{equation}\label{eqn:quantilefunc}
    \mathcal{L}_\tau(\mathbf{X}_{t+1:t+p}, \hat{\mathbf{X}}_{t+1:t+p}) = \sum_{i=1}^{d_1} \sum_{k=1}^{p} \max\left(\tau (\mathbf{X}_{t+k,i} - \hat{\mathbf{X}}_{t+k,i}), (1-\tau)(\hat{\mathbf{X}}_{t+k,i} - \mathbf{X}_{t+k,i})\right),
\end{equation}
where $\mathbf{X}_{t+k}$ are the true target values, and $\hat{\mathbf{X}}_{t+k}$ are the predicted targets from Equation \ref{eqn:themodel}. By minimizing this loss, the model learns to align the predicted quantiles with the empirical distribution of the target variables.

\begin{figure}[htbp]
  \centering
  \includegraphics[width=0.8\textwidth]{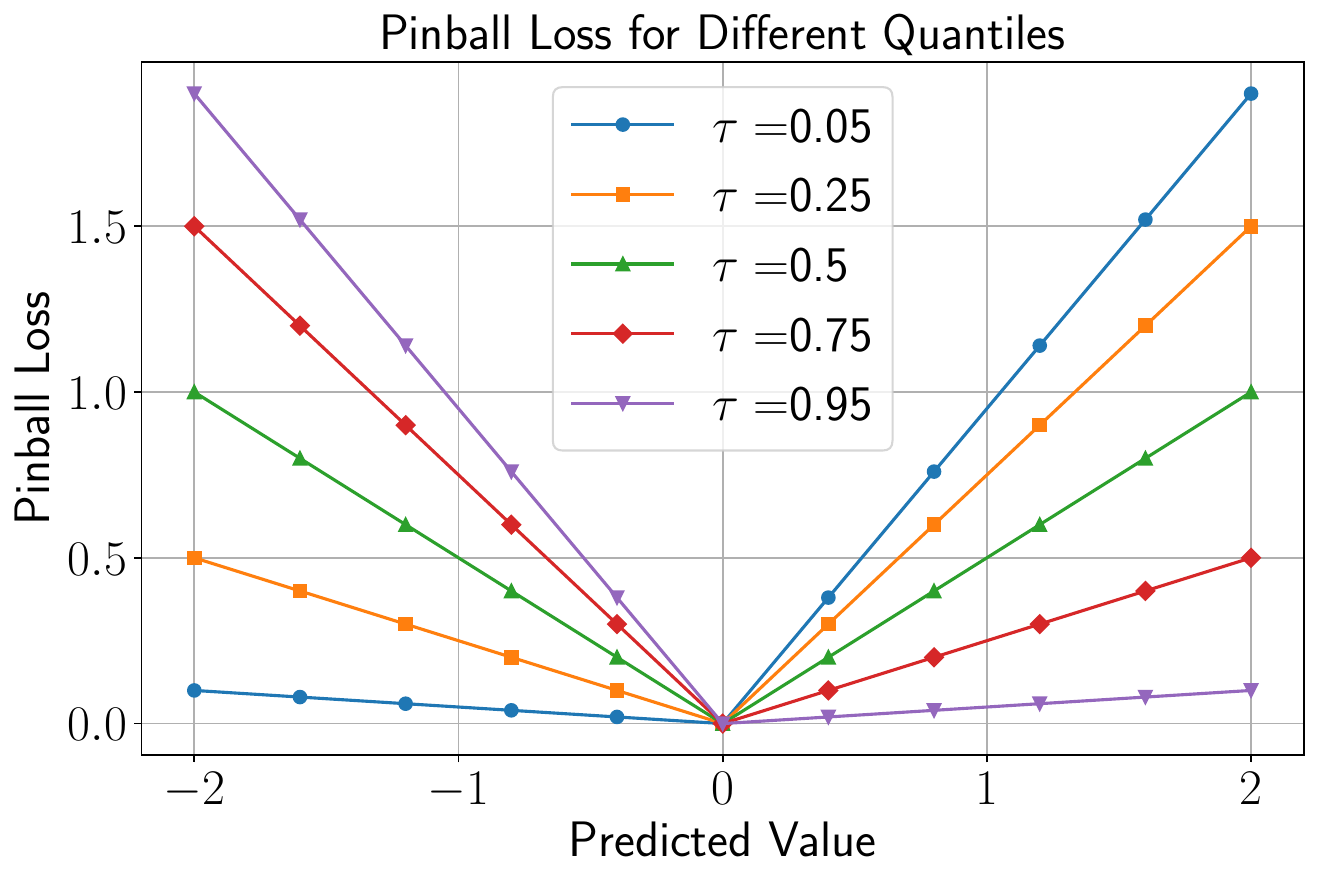}
  \caption[Illustration of the quantile loss function]{An illustration of the quantile loss function for different levels of the desired predicted quantile, $\tau$, 
  given a true target value of $x=0$.
  Each quantile level corresponds to an asymmetric absolute error that penalizes over- or under-prediction differently depending on the quantile $\tau$. When $\tau<0.5$, overestimates incur a larger penalty, encouraging our prediction to be, on average, smaller than the true value; conversely, when $\tau>0.5$ underestimates incur a larger loss penalty. At $\tau=0.5$, the loss reduces to the mean absolute error.}
  \label{fig:quantile_loss}
\end{figure}

This loss function penalizes overestimation and underestimation asymmetrically based on the level of $\tau$, and when $\tau=0.5$ it is equivalent to the mean absolute error, or L1 loss function. 
This loss has been used as a standard forecasting metric for weather models where observations can only be obtained once \cite{james_2000, gneiting_2007, koenker_2017, chung_2021}. Figure \ref{fig:quantile_loss} plots this loss function for several quantile levels, which serves to illustrate the asymmetric penalty induced by quantile levels above or below $\tau=0.5$.

Applying multiple quantile levels to Equation \ref{eqn:quantilefunc}, say, $\mathcal{T}=\{0.05,0.1,0.25,0.5,0.75,0.9,0.95\}$, and averaging, we have
\begin{equation}
    \mathcal{L}_{residual} = \frac{1}{|\mathcal{T}| \ d_{1} \ p} \ \sum_{\tau \in \mathcal{T}}\mathcal{L}_\tau(\mathbf{X}_{t+1:t+p}, \hat{\mathbf{X}}_{t+1:t+p}).
\end{equation}

Finally, incorporating the linear component of the model, the total linear plus residual multi-quantile loss for a given test segment from times $t+1$ to $t+p$, is
\begin{align}
    \mathcal{L}_{\text{total}}\left(\mathbf{X}_{t+1:t+p}, \hat{\mathbf{X}}_{t+1:t+p}^{L}, \hat{\boldsymbol{\epsilon}}_{t+1:t+p}\right) = \mathcal{L}_{\text{linear}} + \mathcal{L}_{\text{residual}}.
\end{align}
Using our seven selected quantile levels, we may generate empirical estimates of the $90^{th}-$percent CI as well as the interquartile range (IQR), defined by the region between the $25^{th}$ and $75^{th}$ percentiles. We explored the use of different quantiles, i.e. $\tau={0.01, 0.02, \dots, 0.99}$, but found that specifying more quantiles did not increase the calibration significantly. The choice here was based only on the desired uncertainty quantification regions. Additionally, using fewer quantiles slightly reduces the number of model parameters and thus increases training speed. This is due to the smaller output dimension of the final feedforward layers in the decoder, Equations \ref{eqn:quantileoutfof2}, \ref{eqn:quantileouthmf2}, and \ref{eqn:quantileoutTEC}.

Conceptually, our transformer network is adding a nonlinear corrective term to the linear predictor by learning to predict the median residual, while also quantifying uncertainty in the forecast in a manner that incorporates all available data from both past and future. It does this without relying on any assumptions of the underlying distributions of the residuals. In practice, we also dynamically weight each component of the loss function, as the loss can become dominated by the errors of the na\"ive linear predictor $\mathcal{L}_{\text{linear}}$. We balance the loss function using learnable weights from each model component, $w_{\text{linear}}$ and $w_{\text{residual}}$, so that
\begin{equation}
    \mathcal{L}_{\text{total}} = e^{-w_{\text{linear}}} \cdot \mathcal{L}_{\text{linear}} + e^{-w_{\text{residual}}} \cdot \mathcal{L}_{\text{residual}} + (w_{\text{linear}} + w_{\text{residual}}).
\end{equation}
These weights adjust the contribution of each loss component during training, and the exponentiation ensures that they are always positive and inversely related to the magnitude of the respective component \cite{kendall_2018}. We found that this weighting reduced the number of epochs needed to reach our early stopping criterion, which was set to halt training when the validation loss was not reduced by at least $1\text{e}^{-4}$ for more than 8 epochs. Additionally, we include dropout layers \cite{hinton_2012} in each of the transformer's multi-head attention blocks to further control overfitting. 

The linear error term only depends on the parameters of each linear model component and thus can be pre-trained using multiple least squares fits. However, this could introduce difficulties if the LIFT model is trained on larger datasets, as the least squares equations would require significant memory. We also found that including the linear components as additional linear layers and training them with Adam did not significantly increase training time. To train the linear and residual components more effectively, we detach the gradients for the linear loss term from the residual loss term. This stops the quantile loss from updating the linear head. It makes the residual used by the quantile term a constant with respect to the linear head, so the linear head is trained only by MSE and the quantile head is trained to model the distribution of those residuals


\subsection{Positional Encoding}\label{sec-pos_encoding}
The basic transformer architecture does not have any inherent mechanism to preserve the temporal or sequential structure of the data. Therefore, it is often necessary to provide a positional encoding feature. The now conventional positional encoding from \cite{vaswani_2017} only gives relative position information for each data segment, so, we also equip our transformer with a time vector. The time vector provides the model with absolute position information, which we split into quadrature components,
\begin{align}
    \mathbf{t}(t) &= 
    \begin{bmatrix}
        \cos\left(\frac{2\pi\text{YOS}_{t}}{11\times365.25}\right) \\
        \cos\left(\frac{2\pi \text{DOY}_{t}}{365.25}\right) \\
        \cos\left(\frac{2\pi \text{HOD}_{t}}{24}\right) \\
        \cos\left(\frac{2\pi \text{MOH}_{t}}{60}\right) \\
        \sin\left(\frac{2\pi \text{YOS}_{t}}{11\times365.25}\right) \\
        \sin\left(\frac{2\pi \text{DOY}_{t}}{365.25}\right) \\
        \sin\left(\frac{2\pi \text{HOD}_{t}}{24}\right) \\
        \sin\left(\frac{2\pi \text{MOH}_{t}}{60}\right)
    \end{bmatrix}^{T},
\end{align}

where $\text{YOS}_{t}$, $\text{DOY}_{t}$, $\text{HOD}_{t}$, and $\text{MOH}_{t}$ are components of the time variable $t$ that represent the year-of-solar-cycle, day-of-year, hour-of-day, and minute-of-hour respectively \cite{poole_2000, athieno_2017}.

\subsection{Model covariates}\label{sec-model_covariates}
\begin{figure}[htbp]
  \centering
  \includegraphics[width=0.9\textwidth]{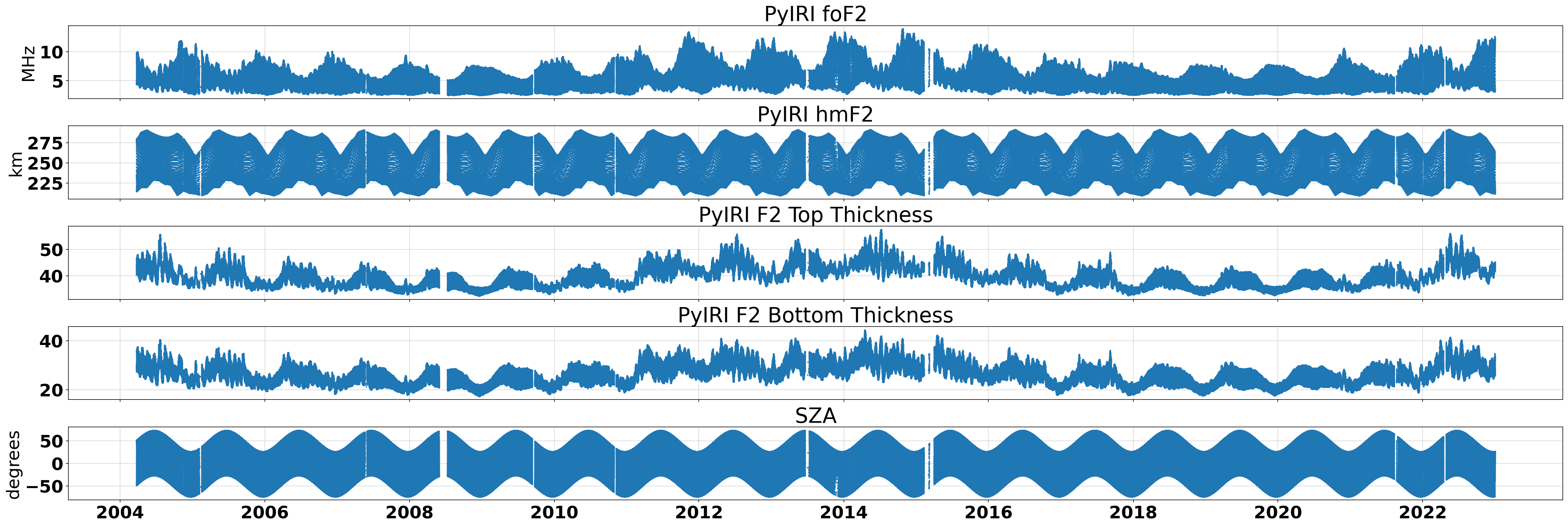}
  \caption[LIFT model covariates]{Model covariates that form $\mathbf{Z}_{t-c:t+p}$, which are computable and provide the encoder and decoder networks with climatology (PyIRI) and solar position data (SZA) for a given sounding station.}
  \label{fig:data_covariates}
\end{figure}

Incorporating additional covariates into the transformer model is as easy as concatenating another time series to the matrix $\mathbf{X}$. Moreover, what sets the transformer apart from most other empirical and assimilative models is its natural ability to incorporate what may be considered future-known covariates through the decoder block (see Figure \ref{fig:model_diagram}). Since PyIRI is a quick and efficient means of generating climatological predictions of the F2-layer parameters, it is an obvious choice for a decoder input. We use PyIRI to generate foF2, hmF2, and the thickness parameters $B_{bot}^{\text{F2}}$ and $B_{top}^{\text{F2}}$ for every contiguous data segment. Together, these PyIRI covariates allow the transformer to include climatology for the F2-region of the ionosphere that uses the latest parameterization of the observed F10.7 cm solar flux. Finally, the solar zenith angle (SZA) is calculated for each geographic location and time \cite{reda_2008}. Figure \ref{fig:data_covariates} illustrates these additional covariates for the Boulder, CO, USA ionosonde station.

\section{Results}\label{sec-results}
\begin{figure}[htbp]
  \centering
  \includegraphics[width=0.65\textwidth]{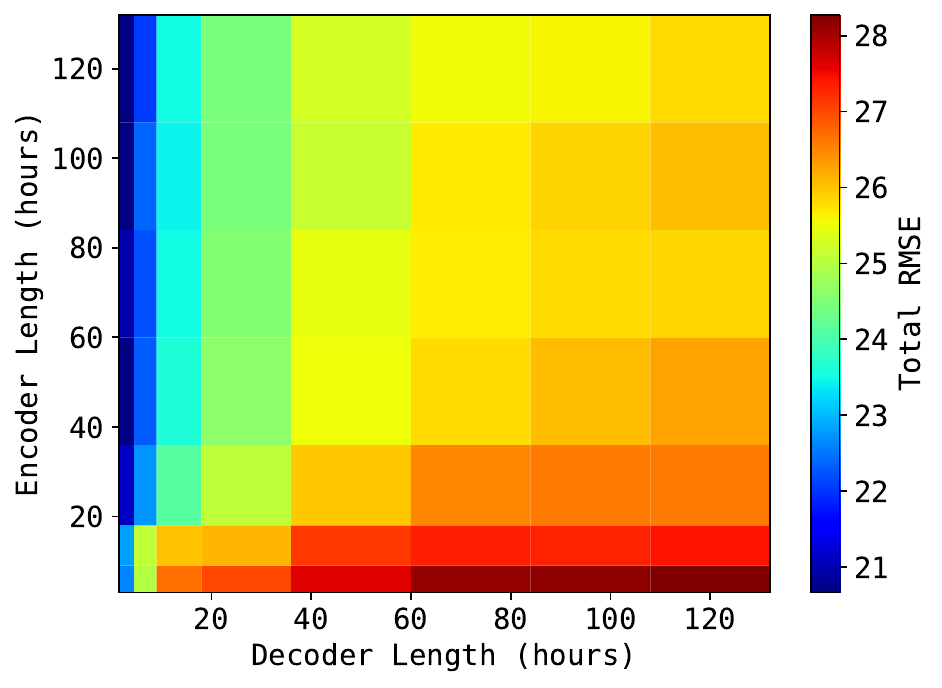}
  \caption[LIFT RMSE for varying encoder and decoder lengths]{Sum of RMSE for foF2, hmF2, and TEC for varying encoder and decoder sequence lengths. The RMSE reported here is the cumulative RMSE over the three forecast parameters for the median prediction.}
  \label{fig:sequence_length_grid}
\end{figure}

A challenge with any time series model is determining the appropriate length of data to provide as input. In classic Box-Jenkins methods, a simple partial autocorrelation can inform the maximum input sequence length, and while previous work \cite{schmolter_2024} suggests that delayed responses of certain ionospheric parameters to solar activity can exceed 48 hours, giving a rough estimate for a desired time lag, our transformer is learning to embed all data in high-dimensional latent variables, thus potentially altering any correlations one may find through the raw data alone. To address this, we adopt a brute-force method and train multiple models with different sequence lengths for the encoder and decoder networks. This was only viable given the relatively reasonable training times of roughly 1-5 hours per model, which we carried out in parallel on a computer with 4 NVIDIA A100 GPUs. Figure \ref{fig:sequence_length_grid} shows the result of this model search in which we find a clear relationship between longer encoder lengths and a reduction in the overall root mean squared error (RMSE).

\begin{figure}[htbp]
  \centering
  \includegraphics[width=0.75\textwidth]{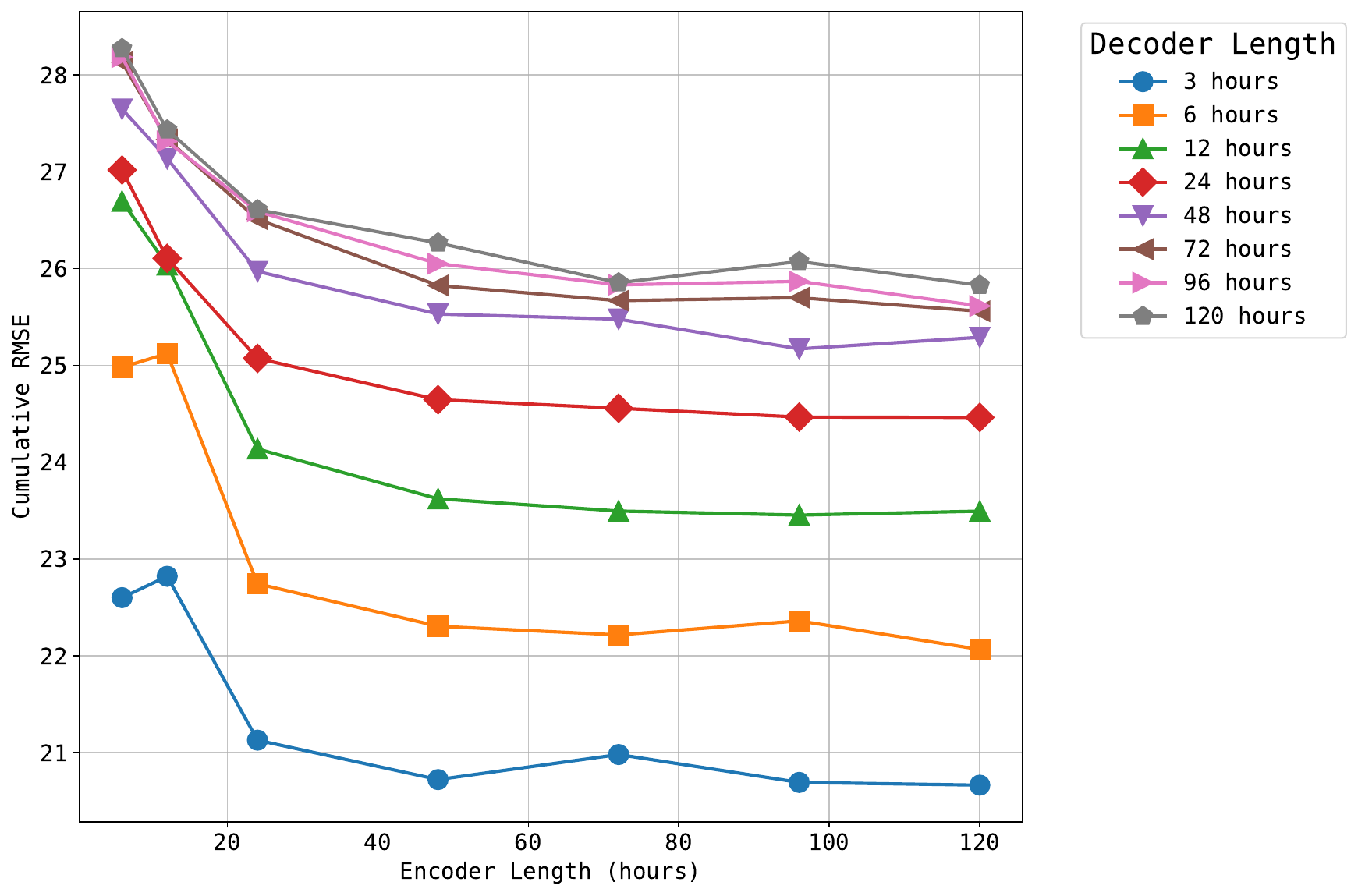}
  \caption[LIFT RMSE for varying encoder lengths]{Sum of RMSE for foF2, hmF2, and TEC for varying encoder lengths for a fixed output length of 24-hours. }
  \label{fig:sequence_length_lines}
\end{figure}

We are primarily interested in the model's ability to produce a 24-hour forecast; Figure \ref{fig:sequence_length_lines} shows the same RMSE values, restricted to a decoder length of 24 hours.
In light of Figures \ref{fig:sequence_length_grid} and \ref{fig:sequence_length_lines}, we chose an input sequence length of 3 days: beyond 3 days, we find that the total RMSE for a 24-hour forecast is only marginally improved with longer input sequences.  This input length balances the accuracy of the forecast against the need for very long sequential segments from a given sounder, which makes the model more practical in an operational setting. In practice, of course, one may retrain this model to fit their specific needs for forecast horizon.

\begin{figure}[htbp]
  \centering
  \includegraphics[width=\textwidth]{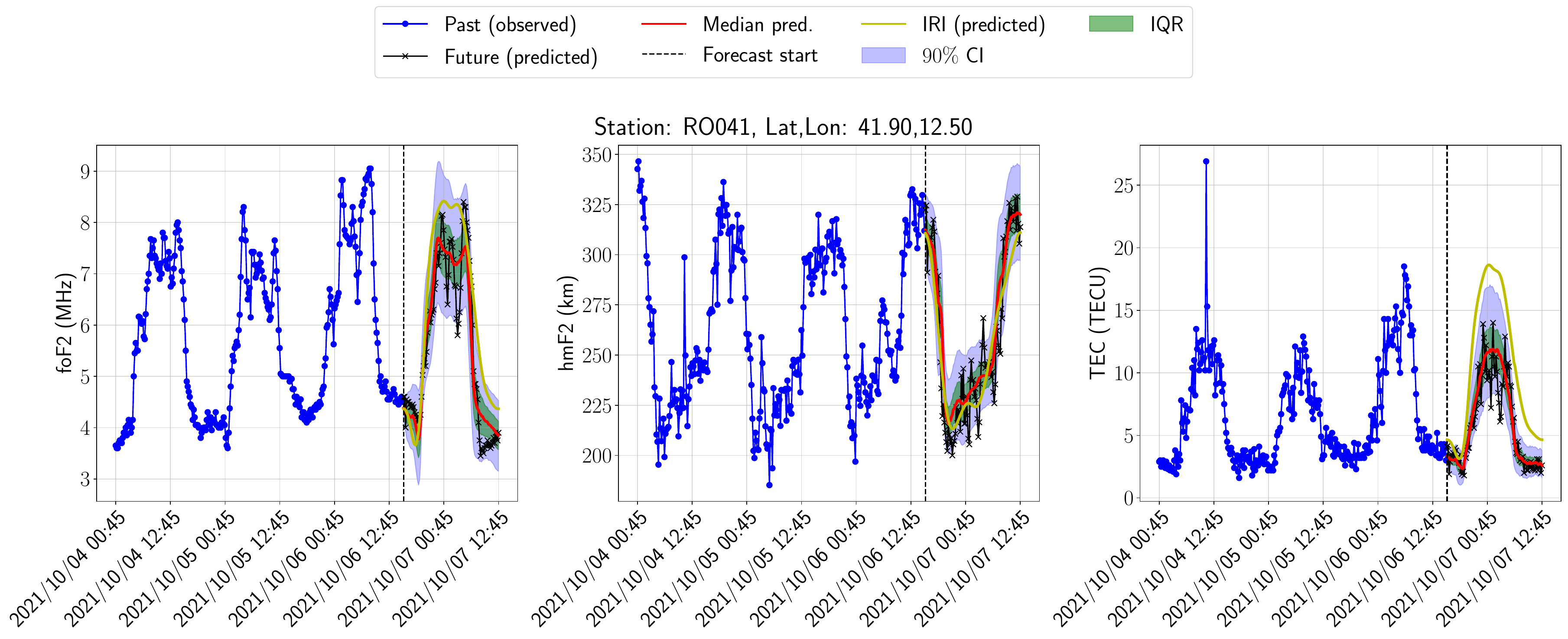}
    \caption[LIFT forecasts for Rome, Italy]{Example 24-hour LIFT forecast for a mid-latitude test station RO041 in Rome, Italy (geographic: $41.90^{\circ}$N, $12.50^{\circ}$E). The left portion of each panel shows the 72-hour input context (past observations), and the right portion shows the 24-hour forecast horizon. The solid red line indicates the LIFT median prediction, shaded regions represent the 90\% confidence interval (blue) and interquartile range (green), and the yellow line shows the PyIRI climatological prediction.}

  \label{fig:test_mid_lat}
\end{figure}

\begin{figure}[htbp]
  \centering
  \includegraphics[width=\textwidth]{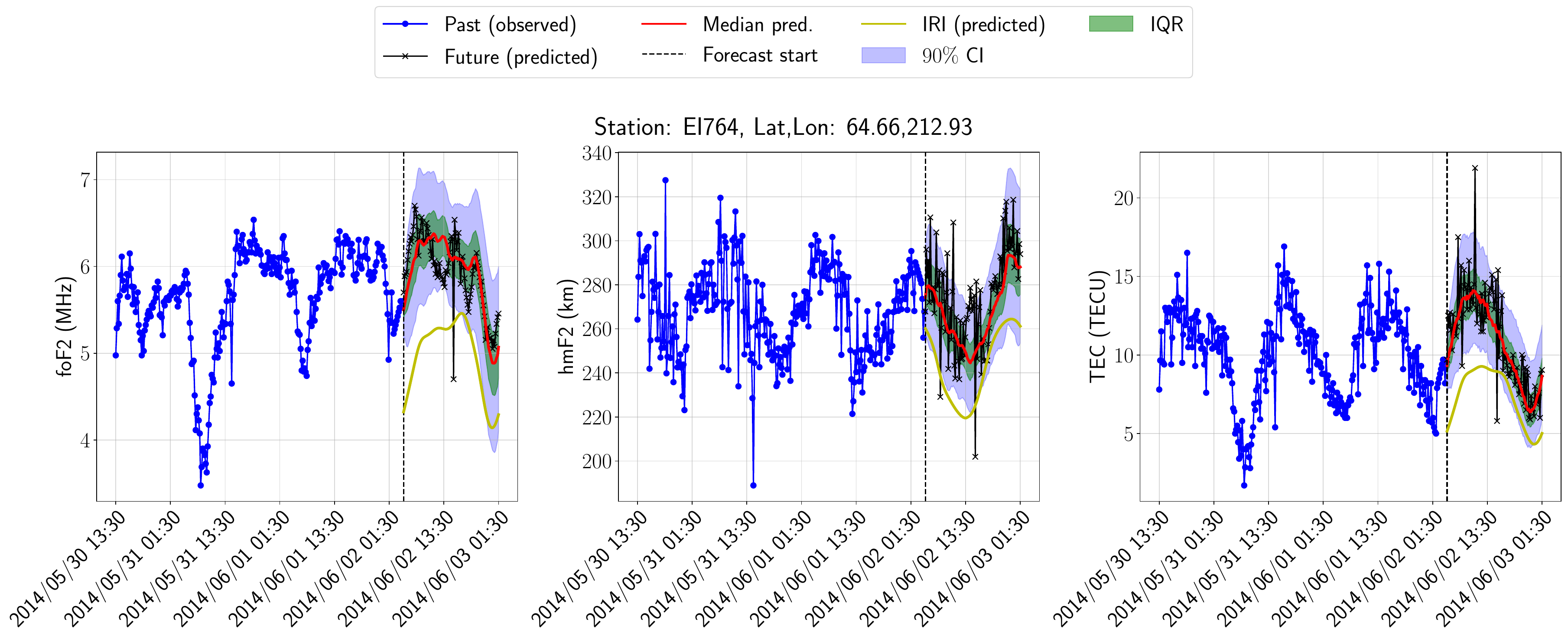}
  \caption[LIFT forecasts for Fairbanks, AK, USA]{Example 24-hour LIFT forecast for a high-latitude test station EI764 near Fairbanks, Alaska, USA (geographic: $64.66^{\circ}$N, $212.93^{\circ}$E). Format follows Figure \ref{fig:test_mid_lat}.}

  \label{fig:test_high_lat}
\end{figure}

\begin{figure}[htbp]
  \centering
  \includegraphics[width=\textwidth]{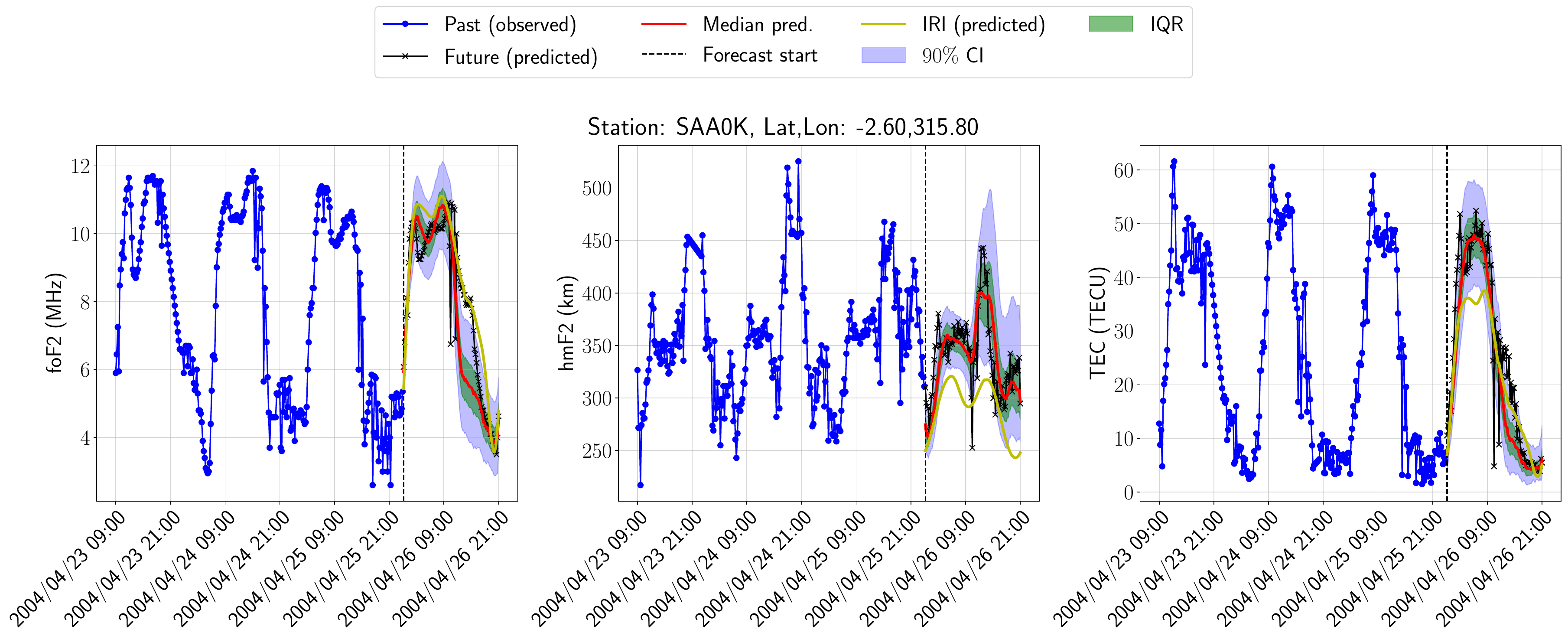}
  \caption[LIFT forecasts for Sao Luis, Brazil]{Example 24-hour LIFT forecast for a low-latitude test station SAA0K in Sao Luis, Brazil (geographic: $2.60^{\circ}$S, $315.80^{\circ}$E). Format follows Figure \ref{fig:test_mid_lat}.}

  \label{fig:test_low_lat}
\end{figure}

The test stations, which were held out entirely during the training process, allow us to examine the model's ability to generalize to new geographic locations. Randomly selected samples from the test set are provided in the following figures to illustrate the prediction of foF2, hmF2, and TEC in practice along with their respective $90\%$ CI and IQR. Figure \ref{fig:test_mid_lat} is from a mid-latitude sounder located at $(41.90 \text{N}, 12.50 \text{E})$, and additional forecasts for high-latitude $(64.66 \text{N}, 212.93 \text{E})$ and low-latitude $(-2.60 \text{N}, 315.80 \text{E})$ stations are given in Figures \ref{fig:test_high_lat} and \ref{fig:test_low_lat}, respectively. These forecasts show, qualitatively, the model's ability to produce uncertainty quantification that is dynamic and dependent on the given segment of data.

\subsection{Forecast accuracy}\label{sec-accuracy}
\begin{figure}[htbp]
  \centering
  \includegraphics[width=\textwidth]{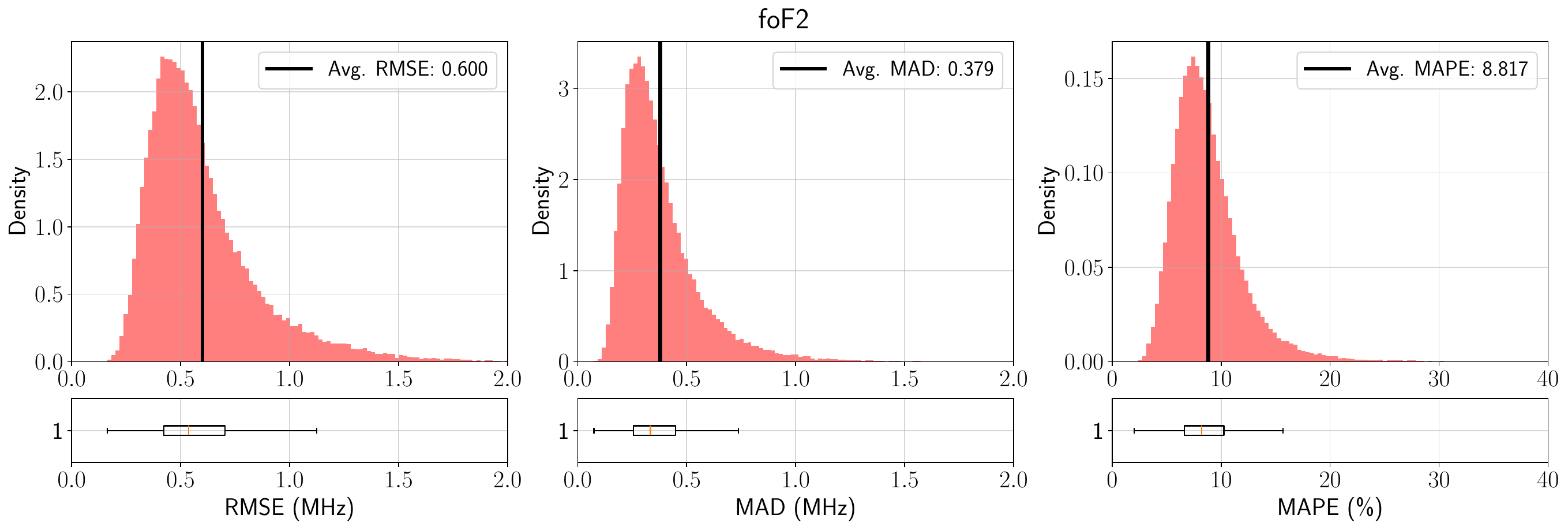}
  \caption[Basic LIFT statistics for foF2 forecasts]{Model forecast RMSE for the foF2 parameter averaged across all forecasted time periods.}
  \label{fig:test_fof2}
\end{figure}

\begin{figure}[htbp]
  \centering
  \includegraphics[width=\textwidth]{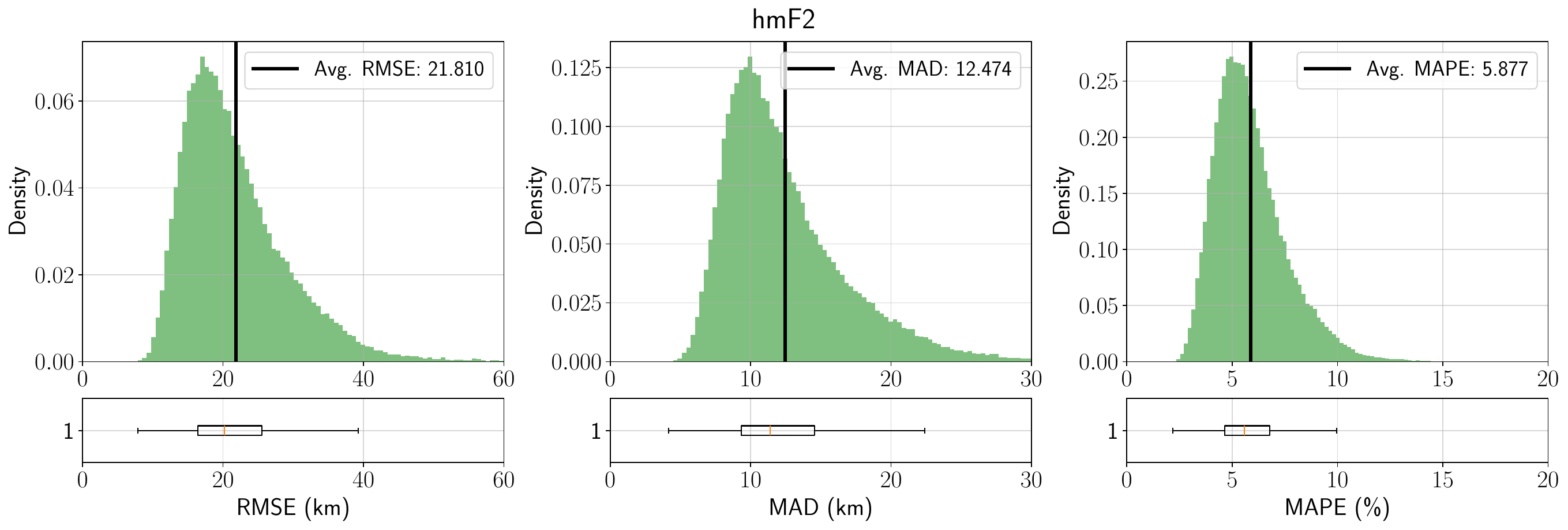}
  \caption[Basic LIFT statistics for hmF2 forecasts]{Model forecast RMSE for the hmF2 parameter averaged across all forecasted time periods.}
  \label{fig:test_hmf2}
\end{figure}

\begin{figure}[htbp]
  \centering
  \includegraphics[width=\textwidth]{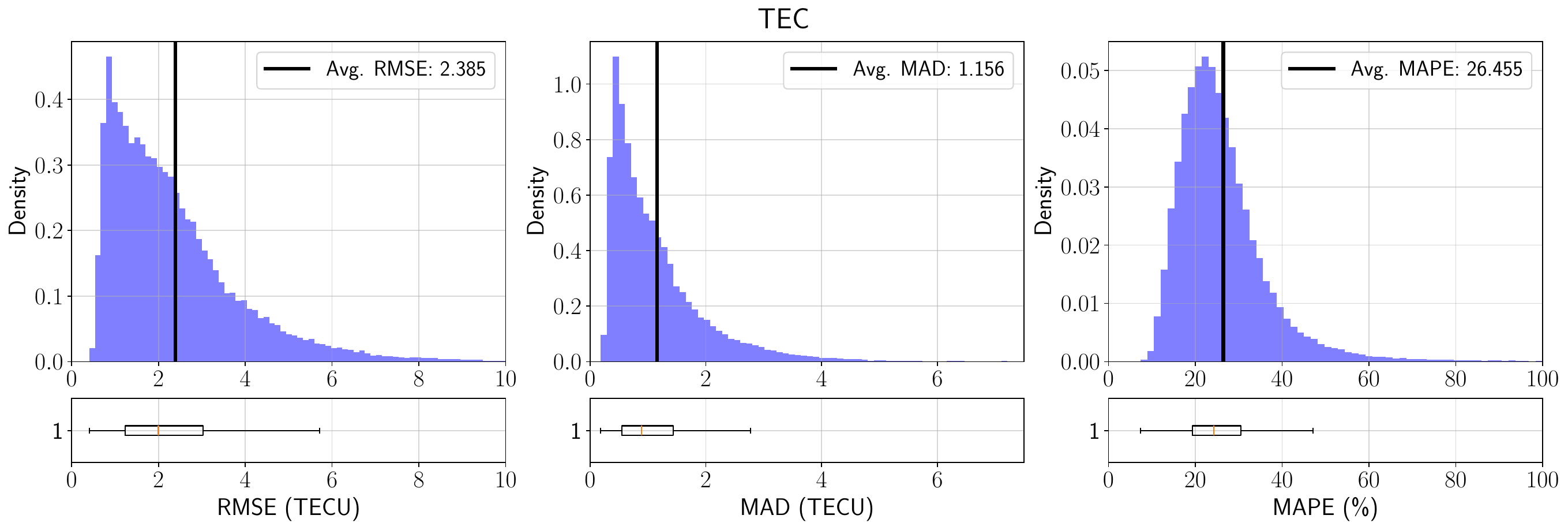}
  \caption[Basic LIFT statistics for TEC forecasts]{Model forecast RMSE for the TEC parameter averaged across all forecasted time periods.}
  \label{fig:test_tec}
\end{figure}

We quantify the overall forecast accuracy of the combined linear model plus median predictions from the transformer using Figures \ref{fig:test_fof2}, \ref{fig:test_hmf2}, and \ref{fig:test_tec}. These figures report the median absolute deviation (MAD) and mean absolute percent error (MAPE) in addition to RMSE. We achieve an overall RMSE of 0.600 MHz for foF2 across all test segments, including high- and low- solar activity and different latitudes. The hmF2 parameter, which is a challenging parameter to predict but has a high impact on many operational systems \cite{bilitza_2022}, has an average RMSE of 21.810 km, and the TEC parameter averages 2.385 TECU over the entire test period. Note that the IQRs for each are represented in the box-plots underneath each error distribution and show that the forecasts, in general, are capturing the point predictions well. These results are consistent with the performance reported in recent machine learning studies: 24-hour TEC forecasts typically achieve RMSE in the range of 2--3 TECU \cite{mao_2025, molina_2025}, and Informer-based foF2 models report RMSE values of 0.5--0.8 MHz depending on forecast horizon and station characteristics \cite{bi2022, qiao2024}. Direct numerical comparison across studies is complicated by differences in test station selection, temporal coverage, and evaluation protocols; nevertheless, LIFT's performance falls within the range of current state-of-the-art methods while additionally providing joint multi-parameter forecasts with calibrated uncertainty quantification.


\begin{figure}[htbp]
  \centering
  \includegraphics[width=\textwidth]{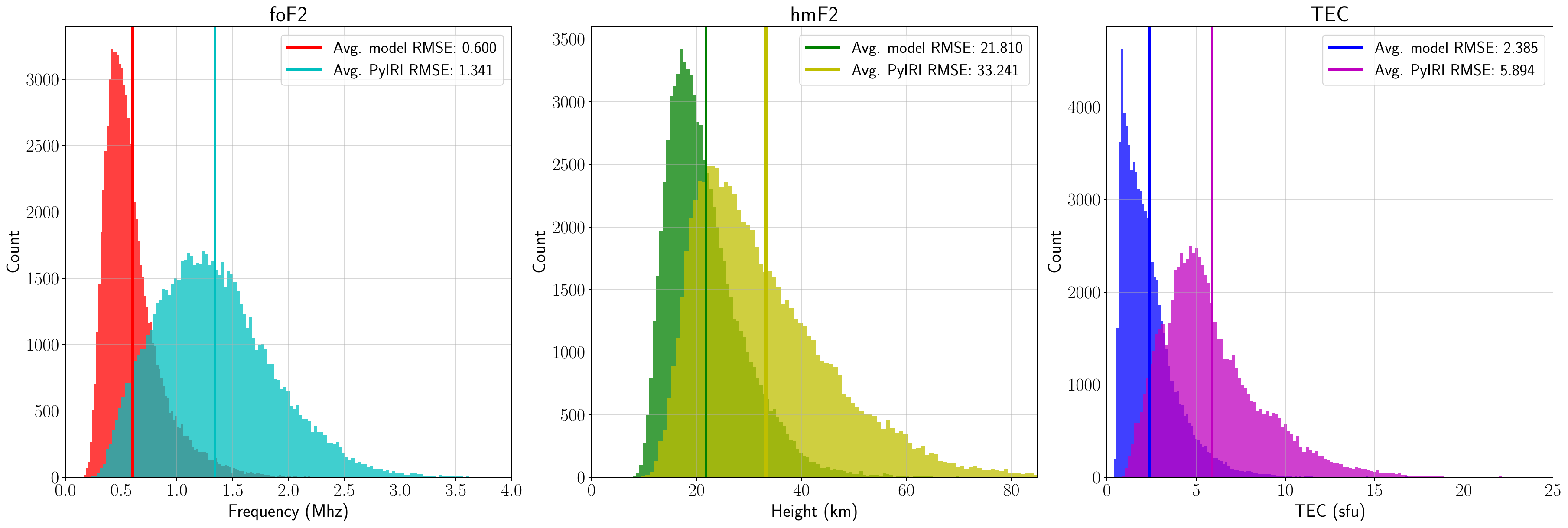}
  \caption[LIFT median forecast performance versus PyIRI]{Model median predictions versus PyIRI parameterized with the F10.7 value at the final context variable time step.}
  \label{fig:model_vs_iri}
\end{figure}

We compare LIFT against PyIRI to quantify the skill improvement achievable by incorporating recent ionospheric observations over climatology alone. Comparisons against IRI-family models are standard practice in the ionospheric forecasting literature, including recent machine learning studies \cite{shih2024, bi2022, qiao2024}, and PyIRI provides a reproducible, well-understood baseline. Direct comparisons with other machine learning models are difficult to conduct in a controlled manner, as most existing transformer-based ionospheric models focus on global TEC maps rather than local ionosonde-derived parameters, train on substantially shorter data records, and do not jointly forecast foF2, hmF2, and TEC with uncertainty quantification \cite{mao_2025}. Comparisons with PyIRI are also straightforward to implement, since it provides the peak height and frequency of the F2-layer as additional covariates to the LIFT encoder and decoder. Figure \ref{fig:model_vs_iri} shows the RMSE error distribution for foF2, hmF2, and TEC using LIFT versus PyIRI. The LIFT model provides an appreciable increase in skill over PyIRI parameterized with the F10.7cm solar flux value from the last observation before the forecast period begins. We also perform a permutation test and generate bootstrap CIs on each set of distributions in Figure \ref{fig:model_vs_iri} to ensure we have statistical significance in the difference of means. P-values and confidence intervals can be see in Table \ref{tab:rmse_stats_test}, all of which show that LIFT's RMSE is statistically significantly lower than PyIRI at an $\alpha=0.05$ level. Despite the best available drivers, PyIRI is still a climatological model, and without computationally expensive data assimilation schemes \cite{bilitza_2017}, the short-term forecast accuracy is difficult to improve. We also include Table \ref{tab:rmse_stats}, which provides a summary of the RMSE performance for the LIFT model as a comparison to a linear-only version of LIFT, a transformer-only version, and the PyIRI forecasts. The illustrates how the transformer and linear baseline complement each other and obtain lower RMSE.

\begin{table}[htbp]
    \centering
    \resizebox{0.75\textwidth}{!}{%
    \begin{tabular}{ccc}
        \hline
        & \multicolumn{2}{c}{Statistical Significance (RMSE mean distributions)} \\ \hline
        Forecast Target & Perm. P-value & Bootstrap 95\% CI \\ \hline
        foF2 (MHz) & p=0.0004 & {[}-0.745, -0.737{]} \\ \hline
        hmF2 (km) & p=0.0004 & {[}-11.543, -11.320{]} \\ \hline
        TEC (TECU) & p=0.0004 & {[}-3.532, -3.485{]} \\ \hline
    \end{tabular}%
}
\caption{Statistical significance for the RMSE predictions for all test data. Permutation test P-values and bootstrap confidence intervals are reported for each of the target forecast variables. Null hypothesis is that the means of the RMSE distributions for LIFT and PyIRI are the same.}
\label{tab:rmse_stats_test}
\end{table}

\begin{table}[htbp]
    \centering
    \small
    \resizebox{0.85\textwidth}{!}{%
    \begin{tabular}{ccccc}
        \hline
        & \multicolumn{4}{c}{RMSE} \\ \hline
        Forecast Target & LIFT & Transformer only & Linear only & PyIRI \\ \hline
        foF2 (MHz) & \textbf{0.600} & 0.898 & 0.637 & 1.341 \\ \hline
        hmF2 (km) & \textbf{21.810} & 32.111 & 22.487 & 33.241 \\ \hline
        TEC (TECU) & \textbf{2.385} & 2.808 & 2.482 & 5.894 \\ \hline
    \end{tabular}%
}
\caption{Summary statistics for the RMSE for all test data. The RMSE is averaged over all points in the test dataset. Comparisons of LIFT to the transformer only and linear only components along with PyIRI. Best performance in RMSE for each target variable is indicated in bold.}
\label{tab:rmse_stats}
\end{table}

\begin{figure}[htbp]
  \centering
  \includegraphics[width=0.75\textwidth]{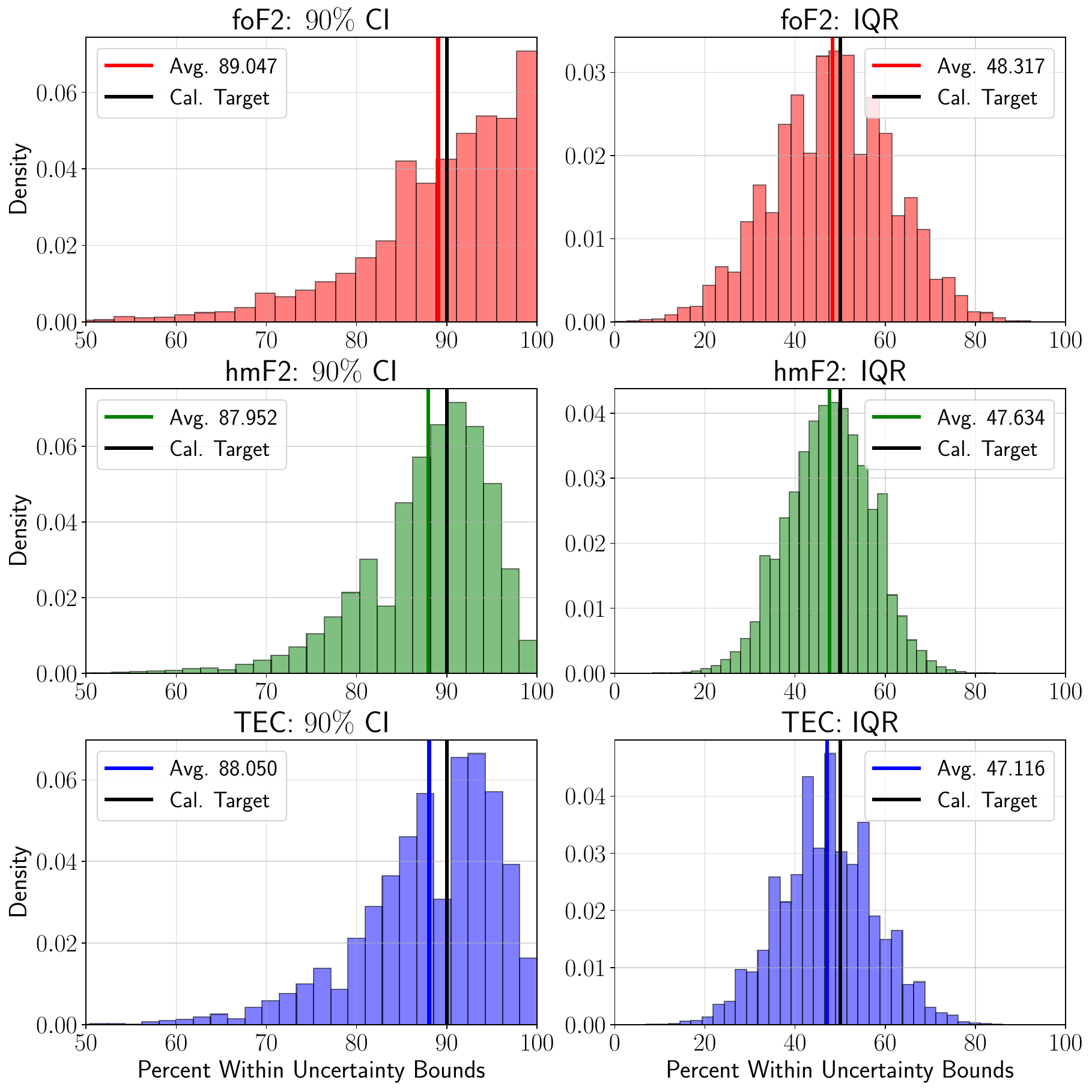}
  \caption[Calibration of the LIFT forecast model]{Model quantile calibration for the foF2, hmF2, and TEC parameters across all test stations for the $90\%$-CI, $\{\tau=0.05, \tau=0.95\}$, and the IQR, $\{\tau=0.25, \tau=0.75\}$.}
  \label{fig:test_quantile_statistics}
\end{figure}

The LIFT model provides uncertainty quantification with each forecast. We evaluate how well calibrated the predicted quantiles are through Figure \ref{fig:test_quantile_statistics}, in which each distribution represents the percentage of points that fall within each confidence region. The statistics for these confidence regions are determined over the full 24-hour duration of each forecast individually,
\begin{align*}
    \text{90\% CI Calibration}(\hat{\mathbf{x}}_{t+1:t+p;j}) &= 100*\frac{1}{p}\sum_{i=1}^{p}\mathbbm{1}(\hat{x}^{0.05}_{t+i;j} < x_{t+i;j}<\hat{x}^{0.95}_{t+i;j}) \\
    \text{IQR Calibration}(\hat{\mathbf{x}}_{t+1:t+p;j}) &= 100*\frac{1}{p}\sum_{i=1}^{p}\mathbbm{1}(\hat{x}^{0.25}_{t+i;j} < x_{t+i;j}<\hat{x}^{0.75}_{t+i;j}),
\end{align*}
where $\mathbbm{1}(condition)$ is the indicator function which equals 1 when \textit{condition} is satisfied and $0$ otherwise. So, for each 24-hour forecast from LIFT, we obtain a single percentage for the number of points correctly within each CI range. Perfectly calibrated quantiles over the entire test dataset would result in the distributions in Figure \ref{fig:test_quantile_statistics} being delta functions at 90 and 50, for the 90\% CI and IQR, respectively. Here we see that both the 90\% CI and IQR are quite well calibrated in terms of their overall mean performance; however, the spread of each distribution implies that uncertainty quantification of LIFT is often too large or too small for many test segments. Table \ref{tab:lowmidhigh-summary} provides a summary of the LIFT model performance at different geomagnetic latitude bands. From this table, we see that the LIFT model maintains reasonably well calibrated uncertainty bands even at the low and high latitudes. We include full histograms of the LIFT quantile calibration stratified by geomagnetic latitude in the Supporting Information. Note, that there are far fewer segments of data to test on at the high latitudes due to the relatively sparse number of GIRO station locations in these regions. This is a limitation of the data available. 

Future iterations of this approach should explore means of better calibrating the uncertainty bounds. Some recent work in this area includes incorporating conformalized quantiles \cite{ramano_2019}, quantile gradient boosting, and Bayesian neural network. While further improvements to the uncertainty calibration are left to future work, our results demonstrate a significant advantage over point forecasts by using a simple linear model for the point predictions and a quantile loss function for the residuals.

\begin{table}[htbp]
\centering
\resizebox{\textwidth}{!}{%
\begin{tabular}{lcccc}
\hline
\multicolumn{1}{c}{} & \begin{tabular}[c]{@{}c@{}}Num. Test \\ Segments\end{tabular} & RMSE & \begin{tabular}[c]{@{}c@{}}Percent Within \\ 90\% CI\end{tabular} & \begin{tabular}[c]{@{}c@{}}Percent Within \\ IQR\end{tabular} \\ \hline
Full Test Set & 76,894 & \begin{tabular}[c]{@{}c@{}}foF2: 0.600\\ hmF2: 21.810\\ TEC: 2.385\end{tabular} & \begin{tabular}[c]{@{}c@{}}foF2: 89.047\\ hmF2: 87.952\\ TEC: 88.050\end{tabular} & \begin{tabular}[c]{@{}c@{}}foF2: 48.317\\ hmF2: 47.634\\ TEC: 47.116\end{tabular} \\ \hline
\begin{tabular}[c]{@{}l@{}}Low Geomag Lat\\ +/-($0^{\circ}$, $20^{\circ}$)\end{tabular} & 24,152 & \begin{tabular}[c]{@{}c@{}}foF2: 0.760\\ hmF2: 26.466\\ TEC: 3.491\end{tabular} & \begin{tabular}[c]{@{}c@{}}foF2: 87.620\\ hmF2: 88.062\\ TEC: 87.115\end{tabular} & \begin{tabular}[c]{@{}c@{}}foF2: 46.727\\ hmF2: 47.833\\ TEC: 46.386\end{tabular} \\ \hline
\begin{tabular}[c]{@{}l@{}}Mid Geomag Lat\\ +/-($20^{\circ}$, $60^{\circ}$)\end{tabular} & 52,598 & \begin{tabular}[c]{@{}c@{}}foF2: 0.528\\ hmF2: 19.687\\ TEC: 1.880\end{tabular} & \begin{tabular}[c]{@{}c@{}}foF2: 89.691\\ hmF2: 87.893\\ TEC: 88.467\end{tabular} & \begin{tabular}[c]{@{}c@{}}foF2: 49.033\\ hmF2: 47.537\\ TEC: 47.441\end{tabular} \\ \hline
\begin{tabular}[c]{@{}l@{}}High Geomag Lat\\ +/-($60^{\circ}$, $90^{\circ}$)\end{tabular} & 144 & \begin{tabular}[c]{@{}c@{}}foF2: 0.319\\ hmF2: 16.267\\ TEC: 1.260\end{tabular} & \begin{tabular}[c]{@{}c@{}}foF2: 93.229\\ hmF2: 90.676\\ TEC: 92.535\end{tabular} & \begin{tabular}[c]{@{}c@{}}foF2: 53.328\\ hmF2: 50.014\\ TEC: 50.666\end{tabular} \\ \hline
\end{tabular}%
}
\caption{Summary of the RMSE and confidence interval (CI) calibration of the LIFT model stratified across low, medium, and high geomagnetic latitude bands. For the $90^{th}$ percentile and the IQR uncertainty bands, perfectly calibrated forecasts would result in 90 and 50 respectively. See Supporting Information for full histograms of the low, mid, and high latitude stratified data.}
\label{tab:lowmidhigh-summary}
\end{table}

\begin{figure}[htbp]
  \centering
  \includegraphics[width=0.97\textwidth]{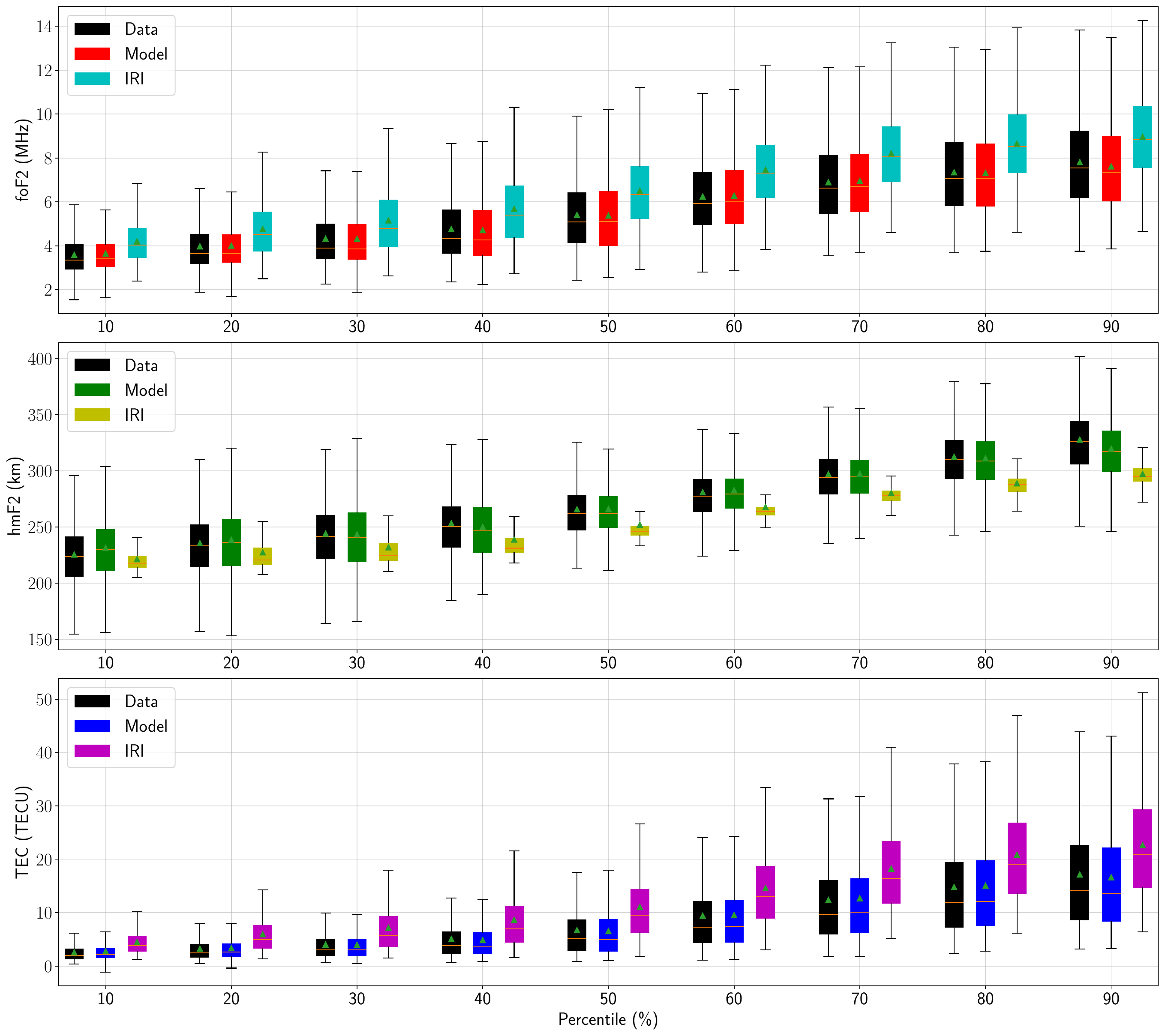}
  \caption[Boxplots for LIFT model bias]{Boxplots of the daily percentile values for (top) foF2 (MHz), (middle) hmF2 (km), and (bottom) total electron content (TEC, in TECU). For each day in the test dataset, the specified percentile (e.g., 10\%, 20\%, …, 90\%) is computed from the observed data (black), the forecast model predictions (colored boxes, as labeled), and the PyIRI climatological predictions (colored boxes, as labeled). Each box therefore represents the distribution across all days of that daily p-th percentile; the box edges show the interquartile range, the horizontal line (or triangle) indicates the median (or mean), and whiskers extend to capture the spread of values. Comparison of the model and PyIRI distributions against the observed distribution provides an overall indication of forecast bias and variability at different percentile levels.}
  \label{fig:test_quantile_boxplots}
\end{figure}
We may also examine our overall model bias as compared to the PyIRI climatological predictions. Figure \ref{fig:test_quantile_boxplots} demonstrates an improvement in the total model bias using LIFT over PyIRI. Each boxplot corresponds to the distribution across all 24‐hour forecasted periods in the test dataset of the p-th percentile for observed data, the forecast model, and the climatological model. That is, for each day we compute, say, the $10^{th}$ percentile of the observed data, and then gather those $10^{th}$‐percentile values across all days into one boxplot---repeating the same procedure for the model and for PyIRI at each percentile. From the figure, we show how the median, interquartile range, and whiskers for that daily percentile varies over the entire test dataset.
While this demonstrates a generally well-calibrated model over the entirety of each 24-hour forecast, further calibration of the model's uncertainty as a function of time-of-day will be a focus for future efforts.

The close visual agreement between model and observed boxplots in Figure \ref{fig:test_quantile_boxplots}, particularly for TEC, reflects successful spatial generalization rather than overfitting. These results are derived from the 76,894 test segments shown in Table \ref{tab:lowmidhigh-summary}, which represent stations completely excluded from training and validation. The distributional alignment demonstrates that LIFT reproduces the statistical properties of ionospheric variability at novel locations, while individual forecasts retain quantifiable error as shown in the RMSE distributions of Figures \ref{fig:test_fof2}--\ref{fig:test_tec}. We note that TEC, being derived from the integrated electron density profile, exhibits a more statistically regular distribution with fewer extreme outliers compared to foF2 and hmF2, which may contribute to the particularly close distributional agreement observed for this parameter.

\subsection{Model Forecast Impacts}\label{sec-impacts}
A common application for improved local ionospheric parameter forecasts is in propagation analysis for long-range radio communications systems. Propagation models are only as good as the environmental specification provided to them, and so in this section we illustrate how the LIFT model can offer improved frequency vs range prediction for high-frequency (HF) radio propagation over PyIRI that includes uncertainty quantification. To do this, we use a numerical ray tracing model that is based on the original Jones Stephenson 3D magnetoionic HF propagation code \cite{jones_1975}. We numerically solve for the ray paths using electron density profiles (EDPs) constructed from LIFT's predicted parameters, specifically, foF2 and hmF2. The ray paths then allow us to determine the distances a radio wave at each frequency will propagate down range. In doing so, we may illustrate how an increase in the prediction accuracy of these parameters and their uncertainty quantification translates directly into measurable improvements for operational users.

\begin{figure}[htbp]
  \centering
  \includegraphics[width=0.75\textwidth]{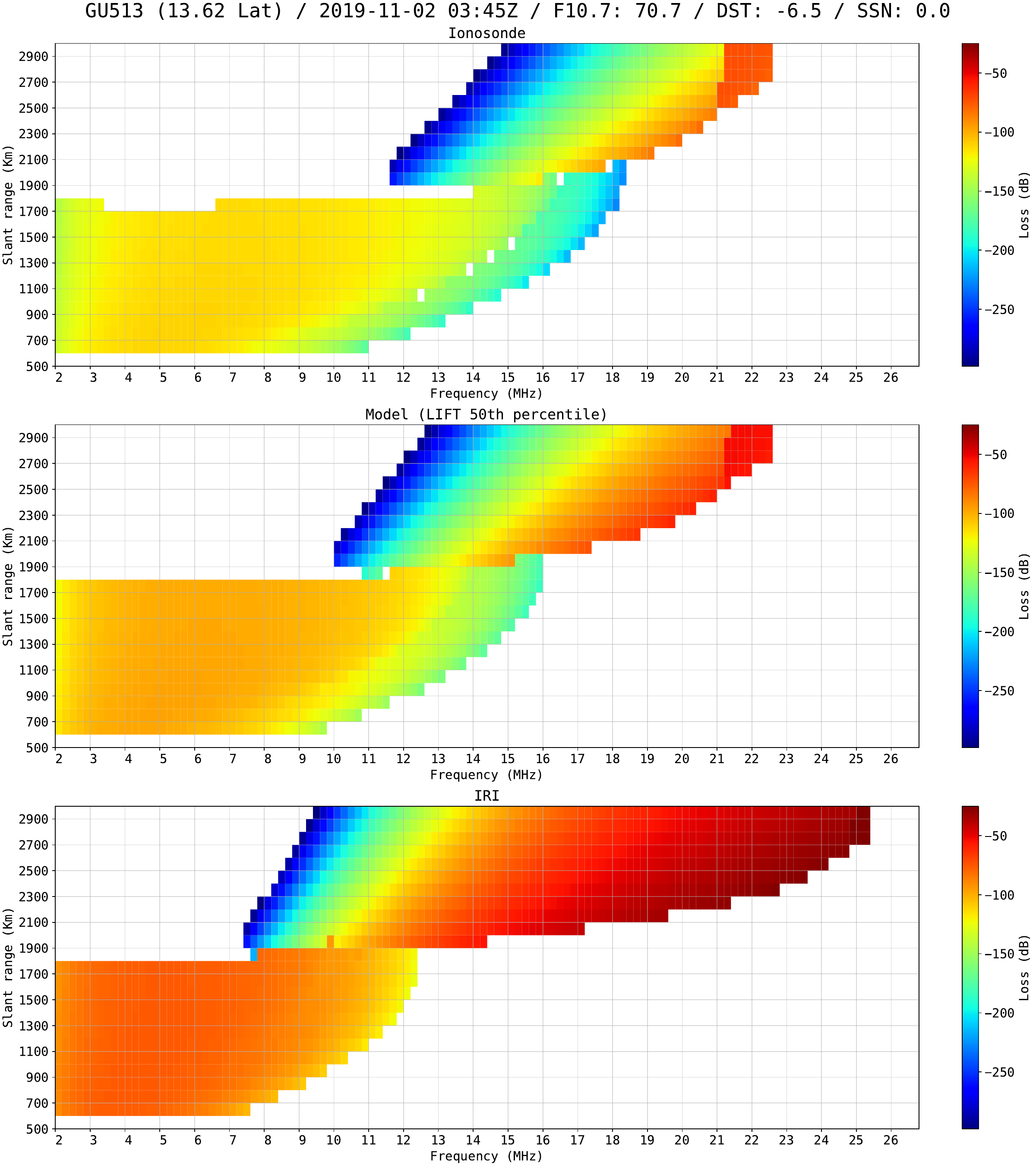}
  \caption[Frequency-range plots derived from LIFT output for low solar activity]{Simulated frequency-range plots using vertical electron density profiles generated from the observed foF2 and hmF2 (True), the LIFT model (Forecasted), and PyIRI predicted (IRI). The date for this simulation was randomly selected but constrained to a period in 2019 of low solar activity.}
  \label{fig:wsbis_low_guam}
\end{figure}

Figure \ref{fig:wsbis_low_guam} shows frequency-range plots for a hypothetical radio transmitter for which the sounding station GU513 (Guam, USA) is roughly at the midpoint of its propagation path. We randomly selected a time from our test period and reconstructed EDPs using parameters from the sounder measurements, our LIFT model, and PyIRI, and then traced rays at frequencies from 3-30MHz through each EDP to give full coverage of the HF radio band. We see in Figure \ref{fig:wsbis_low_guam} that the forecasted frequency-range plot more closely matches that of the true profile. Similar plots are regularly used to help inform the operational space weather community as to what frequencies should be used to propagate out certain distances, wherein the leading edge of these heatmaps shows the maximum usable frequency for a given range. Regarding the leading edge curve generated by the LIFT model's median prediction, it represents a significant improvement over PyIRI, even though PyIRI is being parameterized with the F10.7cm solar flux index from the timestamp of the last input to the encoder module.

\begin{figure}[htbp]
  \centering
  \includegraphics[width=0.75\textwidth]{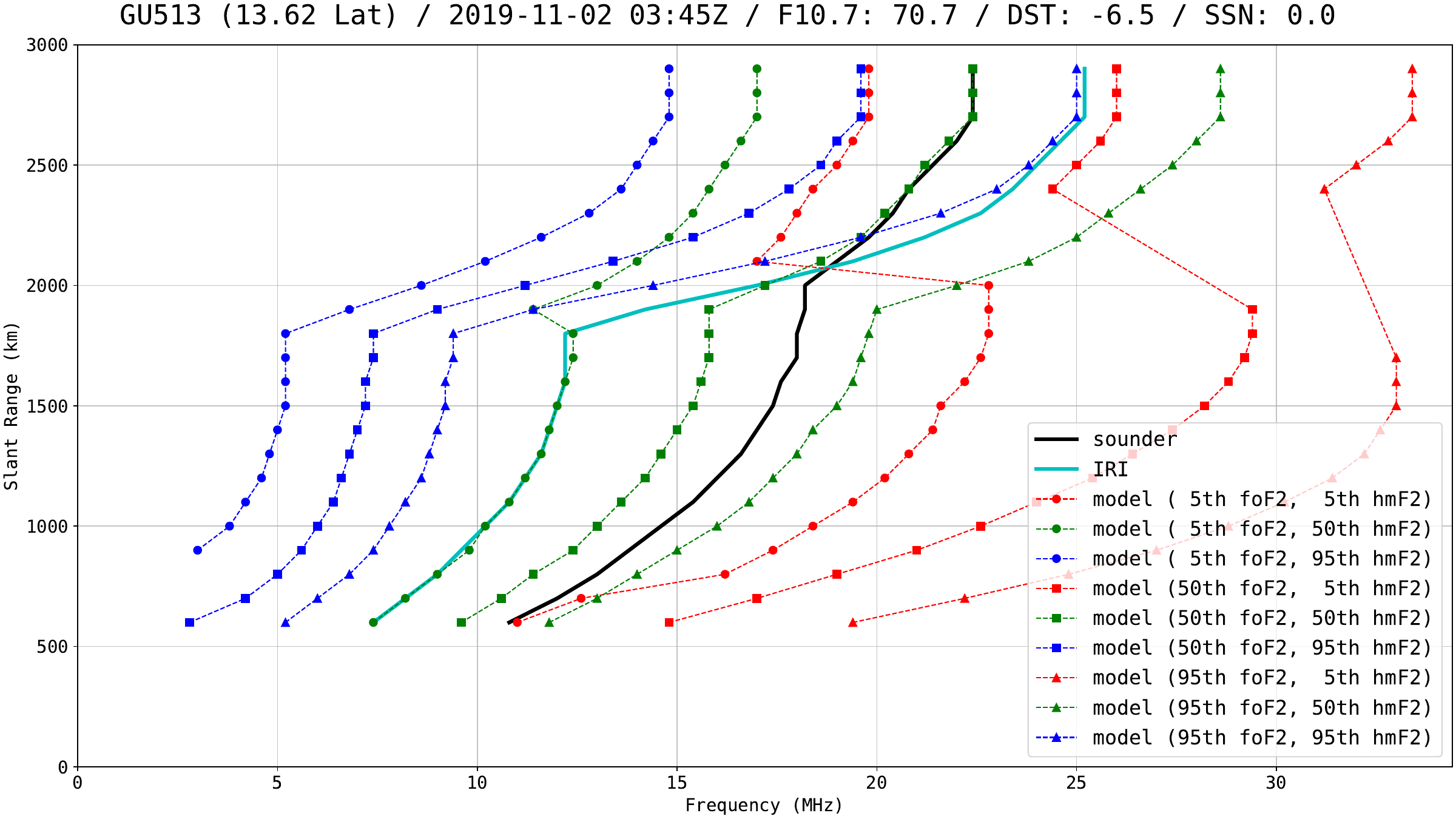}
  \caption[Leading edge plots derived from LIFT output for low solar activity]{Simulated leading edges of frequency-range plots using vertical electron density profiles generated from the observed foF2 and hmF2 (True), PyIRI predicted (IRI), and the upper and lower 5th percentiles from the LIFT model (Forecasted). The date for this simulation was randomly selected but constrained to a period in 2019 of low solar activity.}
  \label{fig:leading_edges_low_guam}
\end{figure}

However, the leading edge curve from the median prediction is only the point forecast. Using the LIFT model CIs for foF2 and hmF2, we may generate additional leading edge curves and produce uncertainty quantification in these downstream propagation tools as well. While Equation \ref{eqn:cipred} represents the $90\%$-CI for both foF2 and hmF2 simultaneously, we may also produce different permutations of the quantile outputs from LIFT for each parameter. Figure \ref{fig:leading_edges_low_guam} illustrates the leading edge curves for these permutations of the estimated LIFT quantiles. These curves reveal the inverse relationship between foF2 and hmF2, where the true leading edge will almost always be bracketed on the left by the combination ($5^{th}$-percentile foF2, $95^{th}$-percentile hmF2) and on the right by ($95^{th}$-percentile foF2, $5^{th}$-percentile hmF2). These represent the extremes in our forecast, and while their purpose remains purely illustrative in this work, they further motivate the need for multivariate uncertainty quantification in future ionospheric forecasting models used for propagation analysis.

\begin{figure}[htbp]
    \centering
    \includegraphics[width=0.95\textwidth]{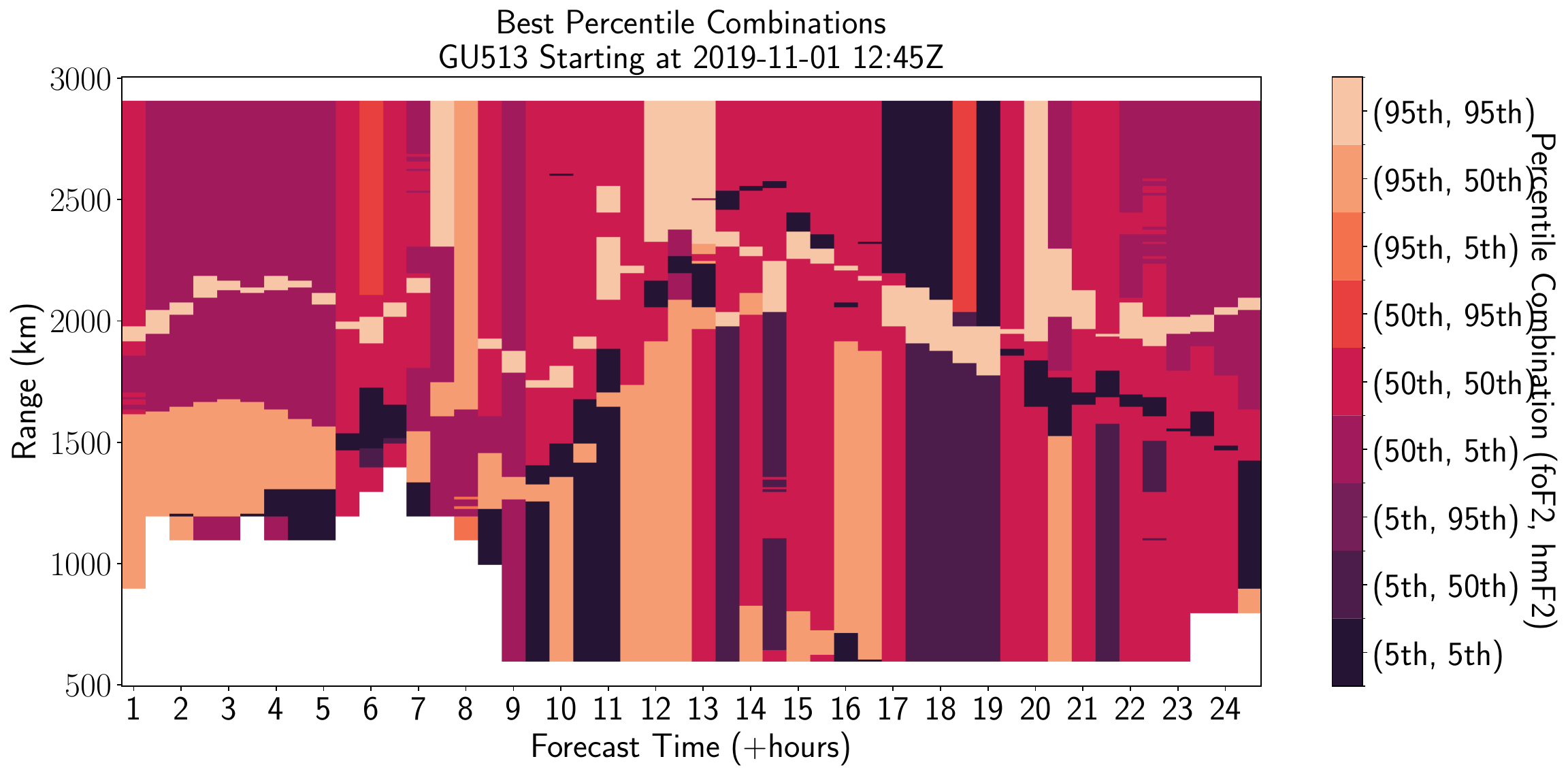}
    \caption[LIFT leading edge performance during low solar activity]{Performance of LIFT quantile forecast in Guam, USA during a period of low solar activity.}
    \label{fig:le_best_q_guam_low}
\end{figure}

\begin{figure}[htbp]
  \centering
  \includegraphics[width=0.95\textwidth]{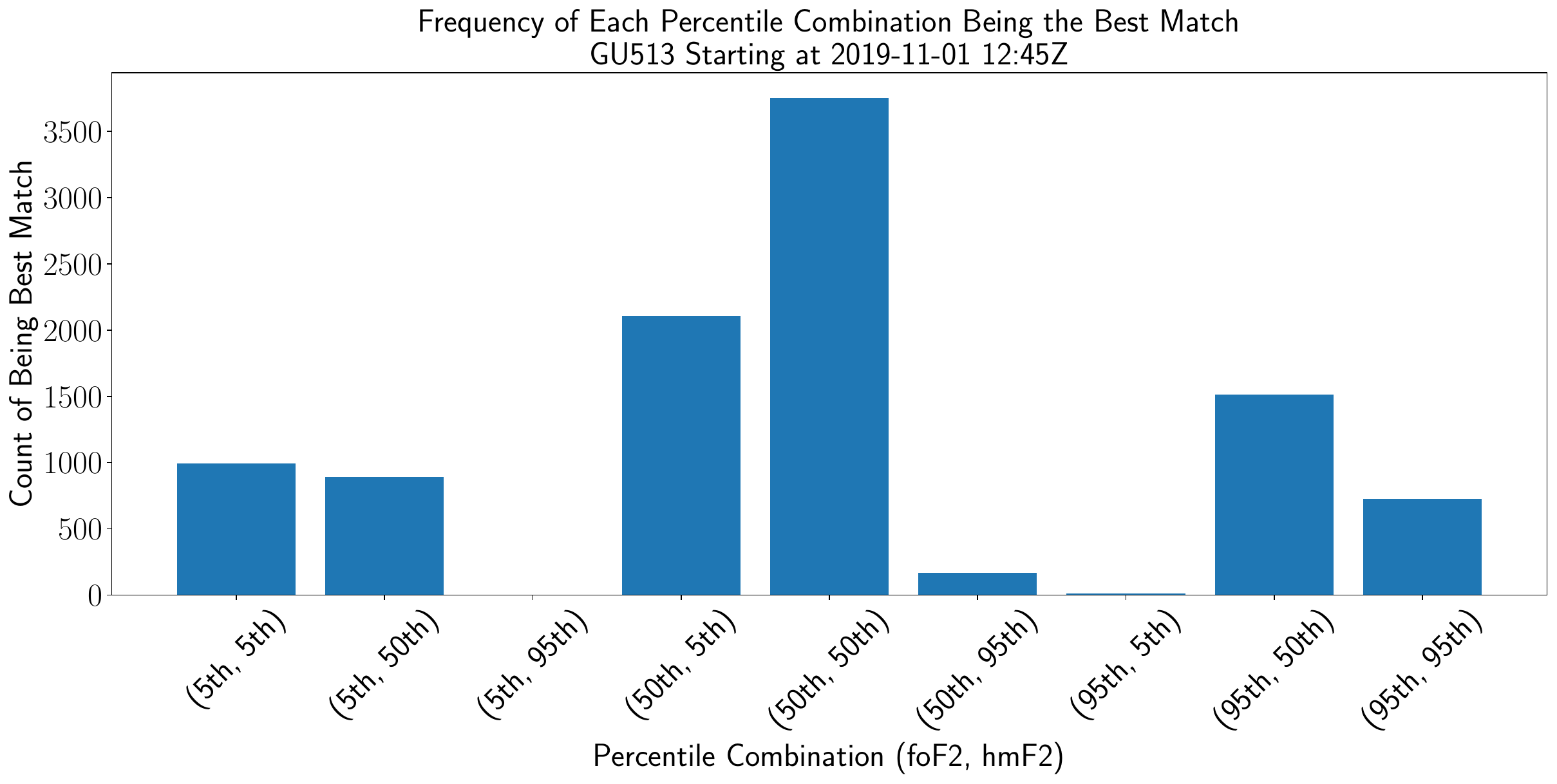}
  \caption[Histogram summary of LIFT leading edge performance for low solar activity]{Histogram reporting which quantile combination of foF2 and hmF2 parameters from the LIFT forecast produced the most accurate HF propagation in Guam, USA during a period of low solar activity.}
  \label{fig:le_hist_guam_low}
\end{figure}

\begin{figure}[htbp]
    \centering
    \includegraphics[width=0.95\textwidth]{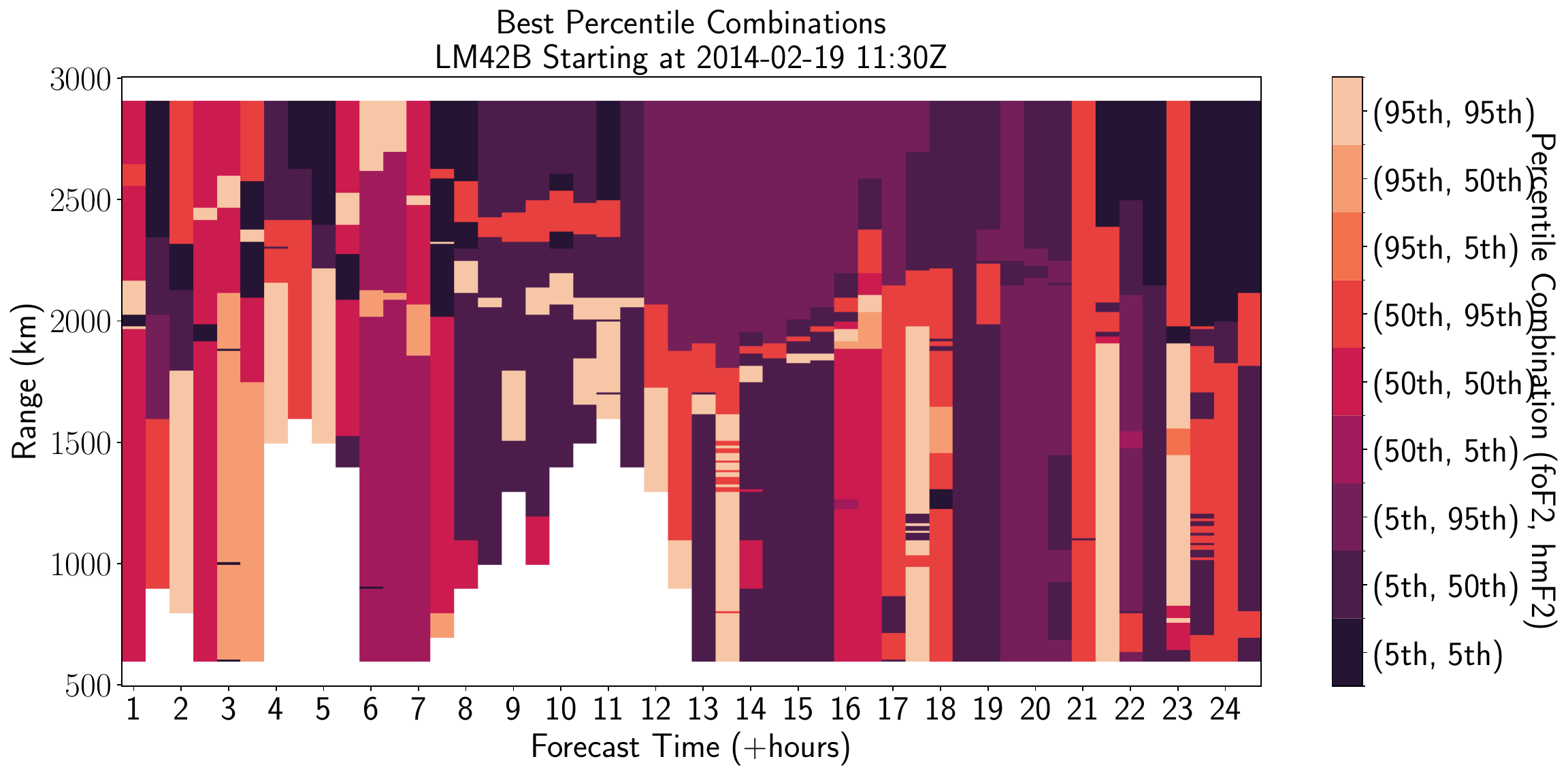}
    \caption[LIFT leading edge performance during high solar activity]{Performance of LIFT quantile forecast in Learmonth, Australia during a period of high solar activity and during a moderate geomagnetic storm (Dst $\le$ -50 nT).}
    \label{fig:le_best_q_lear_high}
\end{figure}

\begin{figure}[htbp]
  \centering
  \includegraphics[width=0.95\textwidth]{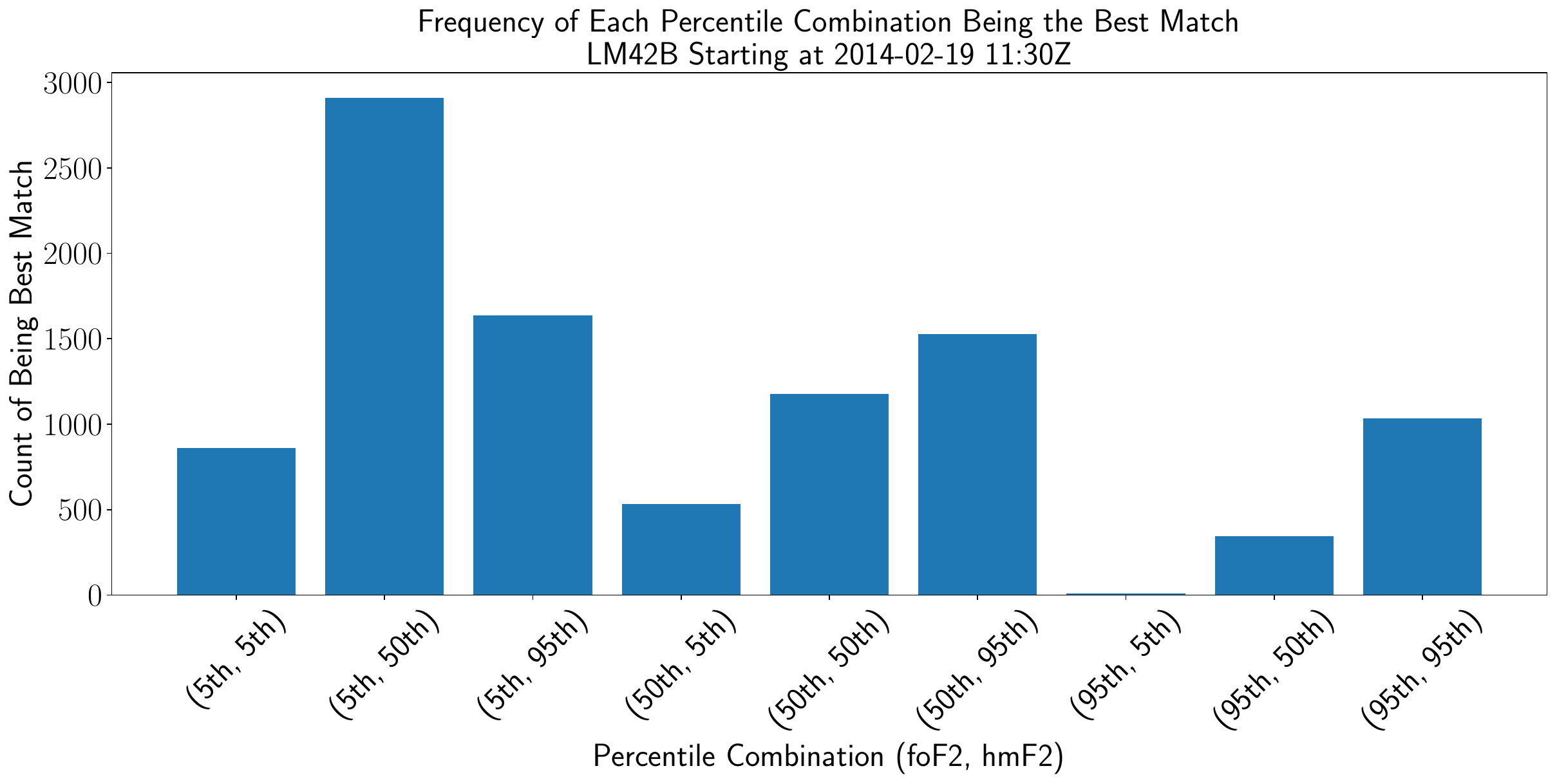}
  \caption[Histogram summary of LIFT leading edge performance for high solar activity]{Histogram reporting which quantile combination of foF2 and hmF2 parameters from the LIFT forecast produced the most accurate HF propagation in Learmonth, Australia during a period of high solar activity and during a moderate geomagnetic storm (Dst $\le$ -50 nT).}
  \label{fig:le_hist_lear_high}
\end{figure}

Figures \ref{fig:wsbis_low_guam} and \ref{fig:leading_edges_low_guam} only illustrate a single time step from the 24-hour LIFT model forecast. By generating the leading edge curves in Figure \ref{fig:leading_edges_low_guam} for every time step in the forecast, we show which combination of quantiles most accurately matched the leading edge derived from the sounder profile as a function of time and range. Figure \ref{fig:le_best_q_guam_low} shows these quantile pairings as a heatmap over the entire 24-hour forecast, and Figure \ref{fig:le_hist_guam_low} provides the histogram for how many range-time bins each combination was the best match of the true leading edge. The histogram indicates that this forecast period was particularly well matched by the median LIFT output ($50^{th}, 50^{th}$). However, Figures \ref{fig:le_best_q_lear_high} and \ref{fig:le_hist_lear_high} demonstrate that the median prediction is not always the best. This particular period in 2014 captures a moderate geomagnetic storm (Dst $\le$ -50 nT). As a result, the LIFT model ($5^{th}, 50^{th}$) quantile prediction more closely matched the true leading edge for a station in Learmonth, Australia.  See Supporting Information for additional frequency-range and leading edge plots during this period. This indicates that, while the median was a poor predictor during the storm, the lower bound of the foF2 forecast was still able to generate a leading edge that covered the observations. While this helps illustrate the importance of uncertainty quantification in ionospheric forecasts, further development of these types of tools will need to be the focus of future work. Improvements to the LIFT model may even seek to incorporate propagation results into the calibration of the model uncertainty. Numerical ray tracing is computationally very expensive, however, it is possible that substituting more efficient approximations such as surrogate models could prove useful in performing model calibration.


\section{Summary and Conclusions}\label{sec-conclusions}
We have presented LIFT, a Local Ionospheric Forecast Transformer that jointly predicts foF2, hmF2, and TEC with calibrated uncertainty quantification. To our knowledge, LIFT is the first transformer-based ionospheric model to combine joint multi-parameter local forecasting, multi-quantile uncertainty bounds, a hybrid linear-transformer architecture, and rigorous spatial generalization testing on geographically held-out stations. The model operates in a pseudo-data assimilation fashion that fuses multiple space weather data sources with na\"ive predictors, treating both the simple linear autoregression and the PyIRI climatology as na\"ive forecasts.

While the transformer makes including exogenous covariates very straightforward, it is also a natural choice for quantile forecasting and uncertainty quantification due to its ability to translate long-range influences in the data into confidence regions that are tailored to each data segment rather than relying on assumptions about the underlying distributions of the errors. The other advantage of transformers for space weather forecasting is their efficiency in processing long time series with relatively high cadence in comparison to conventional recurrent neural networks (RNNs) such as long short-term memory (LSTM) networks. We also establish how the predicted confidence regions for these parameters translate directly into real-world applications for more robust radio propagation assessments and planning through ray traced frequency-range plots.

Although the LIFT model approach addresses several challenges in modern space weather forecasting, it is nevertheless limited to local forecasts and may not provide the most robust predictions under severely disturbed conditions. Future iterations of this approach will therefore focus on extensions to global grids and better calibration of uncertainty, especially in the tails of the distributions and during perturbed conditions. It is important to reiterate that the data used in this study were minimally processed from the original GIRO dataset, only using the ARTIST5 confidence scores to remove poorly autoscaled data. This was intentional and meant to demonstrate that we could train a reasonably well-calibrated system using data as one would expect to find it in real-time from the GIRO repository. This is in contrast to many other data-driven models built from only exquisite, manually scaled EDPs. Finally, transformers require large amounts of data to train on, and time series prediction in particular is much easier when data are contiguous. This is not the case for the other common parameters for the ionospheric profile that specify the F1- and E-regions, but future work is required to include these types of data in the output of this model as they are only intermittently available, i.e. only measurable during certain hours of the day or during certain space weather conditions.

\section*{Data Availability}\label{sec-availability}
The model code is openly available at \url{https://github.com/JayLago/LIFT} and the permanent link to the model and data may be found at \cite{lago2025model}. The raw ionosonde data may be found at \url{https://giro.uml.edu/didbase/}. The raw SSN, Kp, and F10.7 indices were obtained from the GFZ German Research Center for Geosciences, \url{https://kp.gfz.de/en/data}, the direct link to the text file containing the historical record is found at \url{https://kp.gfz.de/app/files/Kp_ap_Ap_SN_F107_since_1932.txt}. Finally, raw Dst data were obtained through the WDC for Geomagnetism, Kyoto \url{http://wdc.kugi.kyoto-u.ac.jp/wdc/Sec3.html}. All of these data are also available in the fully processed dataset included in the permanent link \cite{lago2025model}.

\section*{Conflict of Interest Statement}
The authors have no conflicts of interest to disclose.

\section*{Acknowledgments} \label{sec-acknowledgements}
The authors would like to thank Dr. Victoriya Forsythe for her assistance and input with running the PyIRI model, the code for which can be found at \cite{forsythe_2024}. The authors would like to also thank Dr. Ivan Galkin from the LGDC for the data they continue to collect and make available. We would also like to acknowledge the authors of \cite{zhou_2021, zhou_2023} for the code they made available and which initially formed a basis for constructing our model.

\clearpage
\bibliography{ms}
\bibliographystyle{apacite}
\end{document}